\documentclass[11pt,a4paper]{article}
\usepackage{jheppub}
\pdfoutput=1
\usepackage{hyperref}
\usepackage{graphicx}
\usepackage{amsmath}
\usepackage{amsthm}
\usepackage{amssymb}
\usepackage{tipa}
\usepackage{subfig} 
\usepackage{xcolor}
\usepackage{mathrsfs}
\theoremstyle{plain}

\usepackage{tikz}
\usepackage{pgfplots}
\usepackage{tikz-cd}
\usetikzlibrary{arrows}
\usetikzlibrary{intersections}
\usetikzlibrary{shapes.geometric}
\usetikzlibrary{decorations.pathmorphing, patterns,shapes}

\renewcommand{\bar}{\overline}
\renewcommand{\tilde}{\widetilde}

\renewcommand{\leq}{\leqslant}
\renewcommand{\geq}{\geqslant}
\renewcommand{\Re}{\operatorname{Re}}
\renewcommand{\Im}{\operatorname{Im}}

\newcommand{\Tr}{\operatorname{Tr}}
\newcommand{\nn}{\nonumber}
\newcommand{\Pf}{\operatorname{Pf}}
\newcommand{\const}{\operatorname{const}}
\newcommand{\sgn}{\operatorname{sgn}}
\newcommand{\eff}{\textrm{eff}}
\newcommand{\Diff}{\operatorname{Diff}}

\newcommand{\SU}{\operatorname{SU}}
\newcommand{\SO}{\operatorname{SO}}
\newcommand{\SL}{\operatorname{SL}}

\newcommand{\PSL}{\operatorname{PSL}}
\newcommand{\UU}{\operatorname{U}}

\newcommand{\RR}{\mathbb{R}}

\newcommand{\ZZ}{\mathbb{Z}}

\newcommand{\calD}{\mathcal{D}}
\newcommand{\calF}{\mathcal{F}}

\newcommand{\calT}{\mathcal{T}}

\newcommand{\calJ}{J}

\newcommand{\XLQ}[1]{\textcolor{black}{#1}}
\newcommand{\YG}[1]{\textcolor{black}{#1}}
\newcommand{\DS}[1]{\textcolor{black}{#1}}

\newcommand{\Jeff}{\calJ}

\title{Local criticality, diffusion and chaos in generalized Sachdev-Ye-Kitaev models}

\author[1]{Yingfei Gu,}                                           
\author[1]{Xiao-Liang Qi}                                           
\author[2]{and Douglas Stanford}                                           
\affiliation[1]{Department of Physics, Stanford University, \\
382 Via Pueblo Mall, Stanford, CA, USA}                                           
\affiliation[2]{School of Natural Sciences, Institute for Advanced Study, \\
1 Einstein Dr, 
 Princeton, NJ, USA}                                           
 \emailAdd{yfgu@stanford.edu}                                           
 \emailAdd{xlqi@stanford.edu}                                           
 \emailAdd{stanford@ias.edu}                                          
 \abstract{The Sachdev-Ye-Kitaev model is a $(0+1)$-dimensional model describing Majorana fermions or complex fermions with random interactions. This model has various interesting properties such as approximate local criticality (power law correlation in time), zero temperature entropy, and quantum chaos. In this article, we propose a higher dimensional generalization of the Sachdev-Ye-Kitaev model, which is a lattice model with $N$ Majorana fermions at each site and random interactions between them. Our model can be defined on arbitrary lattices in arbitrary spatial dimensions. In the large $N$ limit, the higher dimensional model preserves many properties of the Sachdev-Ye-Kitaev model such as local criticality in two-point functions, zero temperature entropy and chaos measured by the out-of-time-ordered correlation functions. In addition, we obtain new properties unique to higher dimensions such as diffusive energy transport and a ``butterfly velocity" describing the propagation of chaos in space. We mainly present results for a $(1+1)$-dimensional example, and discuss the general case near the end.
}                                                                                 \arxivnumber{1609.07832}                                           
 
\begin{document}                                          
\maketitle

\section{Introduction}
Holographic duality, also known as the anti-de-Sitter space/conformal field theory (AdS/CFT) correspondence, refers to the duality between quantum many-body systems in $d$ spatial dimensions and gravitational theories in $(d+1)$-dimensional asymptoticaly anti-de Sitter geometries \cite{maldacena1999large,witten1998anti,gubser1998gauge}. Various pieces of evidence suggest that holographic duality is a generic phenomenon that applies beyond the super Yang-Mills theories where the original conjecture was proposed. However, it is difficult to find concrete models for which the duality can be verified directly by comparing bulk and boundary calculations. A very interesting development in this direction is the Sachdev-Ye-Kitaev model\cite{kitaev2015simple}, which is a $(0+1)$-dimensional model describing random interactions between $N$ Majorana fermions. When $N$ is large and the temperature is low, the model has an emergent approximate time reparameterization symmetry which is weakly broken. In this limit, two-point functions and four-point functions can be computed. The results suggest that the model has a weakly coupled holographic dual, which includes dilaton gravity in an approximate AdS$_2$ geometry\cite{Jackiw:1984je,Teitelboim:1983ux,almheiri2014models,
jensen2016chaos,maldacena2016conformal,
engelsoy2016investigation}, weakly coupled to an infinite number of matter fields\cite{maldacena2016comments}.

The SYK model  
is a modification of a quantum spin model proposed by Sachdev and Ye more than 20 years ago\cite{sachdev1993gapless} (which was also related to holographic duality in Ref.~\cite{sachdev2010holographic}). In the large $N$ and low temperature limit $ N \gg \beta \Jeff\gg 1$ (with $\beta$ the inverse temperature and $\Jeff$ the average coupling strength), the behavior of the model is controlled by a large $N$ saddle point, with the fermion two-point function $G(\tau,\tau')=\frac1N\sum_j\langle \chi_j(\tau)\chi_j(\tau')\rangle$ playing the role of a semi-classical ``order parameter" with small fluctuations suppressed by $\frac1N$. At low temperature, $G(\tau,\tau')$ has a power law dependence on $\tau-\tau'$ in the infared, suggesting that the low energy dynamics of this model might be conformally invariant \cite{parcollet1999non}. In fact, the $(0+1)$-d conformal symmetry is only approximate, and in fact the low-temperature dynamics are dominated by the specific way in which the symmetry is broken \cite{kitaev2015simple,maldacena2016comments}.

A particularly interesting type of four-point function is the out-of-time-order correlation function (OTOC) 
$F(t)=\langle \chi_j(t)\chi_k(0)\chi_j(t)\chi_k(0)\rangle$,  which has been proposed as a measure of chaos in quantum systems\cite{larkin1969quasiclassical,shenker2013black,Shenker:2013yza,kitaev2014hidden,roberts2015diagnosing,
shenker2015stringy} (see Ref.~\cite{swingle2016measuring, zhu2016measurement, yao2016interferometric,li2016measuring} for experimental proposals for measuring OTOC). 
Physically, the decrease of this four-point function measures the increase of the size\footnote{At infinite temperature, the ``size" of the anticommutator simply means its 2-norm. At finite temperature, it is generalized to the thermal expectation value $\langle \left\{\chi_j(t),\chi_k(0)\right\}^2\rangle_\beta$. Note that we study the anticommutator instead of the commutator because the operators are fermionic.} of the operator anticommutator $\left\{\chi_j(t),\chi_k(0)\right\}$, which indicates how sensitive the system is to an initial perturbation created by acting with the fermion operator $\chi_j(0)$. The exponential time dependence of the connected part of the OTOC ${F}(t)\YG{_{\rm conn.}}\propto e^{\lambda\YG{_L} t}$ defines an inverse time scale $\lambda\YG{_L}$ which can be considered a quantum analog of Lyapunov exponent. Ref. \cite{maldacena2015bound} proved a general upper bound $\lambda\YG{_L} \leq \frac{2\pi}\beta$ (for a regularized form of OTOC with imaginary time evolution), which is saturated for theories with an Einstein gravity dual. Interestingly, the SYK model also saturates this upper bound
\cite{kitaev2015simple,
polchinski2016spectrum,
maldacena2016comments}.
Many other aspects of the SYK model (and a similar model for complex fermions) have been investigated recently\cite{sachdev2015bekenstein,
polchinski2016spectrum,
you2016sachdev,
fu2016numerical,
anninos2016disordered,
jevicki2016bi,
jensen2016chaos,
engelsoy2016investigation,
Bagrets:2016cdf}.

Given the interesting properties of the $(0+1)$-dimensional SYK model, it is natural to look for higher dimensional cousins. In this paper, we propose a family of higher dimensional variants of the SYK model, which remain solvable in the large $N$ limit. Our model can be defined on an arbitrary discrete lattice in arbitrary spatial dimensions. There are $N$ fermions on each site with an SYK Hamiltonian. Different sites have independent SYK couplings, and neighboring sites are coupled by random four-fermion terms. Using the same techniques as those applied to the SYK model, we can study two-point functions and four-point functions in space-time. The spatial locality of the model allows us to study transport properties and propagation of quantum chaos in space-time. We find that the disorder-averaged two-point functions vanish between different lattice sites, and have the same local critical behavior as in the SYK model within each site. Our model also has the same zero temperature entropy at each site as the SYK model. Correlation between different sites begins at the level of four-point functions. Similar to the SYK model case, one can consider the fermion four-point function as being mediated by a series of collective fields, with the leading contribution coming from energy fluctuations. In our generalized models, the four-point function allows us to study the dynamics of collective fields in space-time. In particular, we find a diffusive dynamics of energy density, which means this model describes a diffusive strongly correlated metal phase. The OTOC can also be studied, which now has both spatial and temporal dependence. We show that at low temperature the Lyapunov exponent still saturates the chaos bound $\frac{2\pi}\beta$. The propagation of quantum chaos in space-time can be characterized by a ``butterfly effect velocity" $x=v_Bt$, which means that the anticommutator $\left\{\chi_{jx}(t),\chi_{jy}(0)\right\}$ becomes significant at $t=|x-y|/v_B + (\const)$. Interestingly, the diffusion constant $D$ and butterfly velocity $v_B$ in our generalized SYK model satisfy a simple relation\footnote{See Ref.~\cite{aleiner2016microscopic,swingle2016slow} for other discussions on diffusion and butterfly effect in solid state systems.
} $D=\frac{v_B^2}{2\pi T}$,
which realises a bound conjectured on diffusion in incoherent metal and agrees with the holographic calculation on incoherent black hole \cite{hartnoll2015theory,
blake2016universalcharge,
blake2016universal}.

The remainder of the paper is organized as follows. In section \ref{sec: review of SYK} we will briefly review the properties of the original SYK model. In section~\ref{section: syk chain model}, we define the generalized SYK model in higher dimensions. For concreteness, we work on a $(1+1)$-dimensional example and study its correlation functions and thermodynamic properties in detail. In section~\ref{section: transport} we study the dynamics of collective fields in this system based on an operator-product expansion of fermion four-point functions. In particular, we show that one of the collective fields describes the diffusion of energy in this system. In section~\ref{section: chaos and the butterfly velocity}, we study the OTOC of the $(1+1)$-dimensional model and obtain the Lyapunov exponent and the butterfly velocity. In section \ref{section: general models at higher dimensions} we discuss the general form of the model in generic dimensions and graphs. In section \ref{section: conclusion and discussion} we end the paper with a summary and discussion of further topics.

\section{Review of the SYK model}
\label{sec: review of SYK}

In this section, we briefly review some basic facts about the SYK model. The SYK model consists of $N$ Majorana fermions ${\chi}_j$, $j=1,2,\ldots, N$ with a random four-fermion interaction \cite{kitaev2015simple}
\begin{equation}
{H}=  \sum_{1\leq j<k<l<m\leq N} J_{jklm} {\chi}_j {\chi}_k {\chi}_l {\chi}_m, \quad \{ {\chi}_j,{\chi}_k \} = \delta_{jk}
\end{equation}
where $\{J_{jklm}\}$  are independent random couplings with zero mean 
$\bar{J_{jklm}}=0$, and variance of individual elements is given by $\frac{1}{3!} N^3 \bar{J^2_{jklm}} = \calJ^2$. Here, the symbol $J$ without indices is a dimensionful constant that sets the scale of the Hamiltonian, and the factors $\frac{1}{3!}N^3$ are for later convenience. The interaction is all-to-all, so that there is no spatial locality, and this model should be considered as a $(0+1)$-d quantum mechanical system. The model is solvable at large $N$, and exhibits holographic behavior at strong coupling $N\gg \beta \calJ\gg 1$. 
In this limit, the (finite temperature) two-point function\cite{sachdev1993gapless,parcollet1999non} $G\left( \tau_1,\tau_2 \right):= \frac{1}{N}\sum_{j=1}^N  \langle \calT_\tau \chi_j(\tau_1) \chi_j(\tau_2) \rangle_\beta$ has emergent $\PSL_2(\RR)$ symmetry\footnote{This symmetry acts as $f\rightarrow \frac{af +b}{cf+d}$ where $f = \tan\frac{\pi \tau}{\beta}$, \YG{and $\begin{pmatrix}
a & b \\
c & d
\end{pmatrix}
\in \SL_2(\RR)$.
}}
\begin{align}\label{conformalSaddle}
G\left( \tau_1,\tau_2 \right)
=&~ b^{\Delta}  \left(  \frac{\beta\calJ}{\pi} \sin \frac{\pi \YG{\tau_{12}}}{\beta} \right)^{-2\Delta},\quad 0\leq \YG{\tau_{12}} < \beta \\
  b=&~ \frac{1}{\pi } \left(\frac{1}{2}-\Delta \right) \tan (\pi \Delta),\quad  \Delta=1/4 \nn
\end{align}
\YG{where we denote $\tau_1-\tau_2$ by $\tau_{12}$ for simplicity here and below}. 
One can also compute the four-point function in the holographic limit\cite{kitaev2015simple, maldacena2016comments}. The interesting piece of the four-point function is the connected part, which begins at order $\frac{1}{N}$
\begin{equation}\label{fourPtDef}\frac{1}{N}\calF (\tau_1,\tau_2,\tau_3,\tau_4):= \frac{1}{N^2}\sum_{j,k=1}^N  \langle \calT_\tau \chi_j(\tau_1) \chi_j(\tau_2) \chi_k(\tau_3) \chi_k(\tau_4) \rangle_\beta - G(\tau_1,\tau_2)G (\tau_3,\tau_4).\end{equation}
At leading order of $1/\beta\calJ$ and $1/N$, when we consider a configuration of imaginary (Euclidean) times with the ordering $\tau_1>\tau_2> \tau_3 > \tau_4$, then the correlator $\calF$ factorizes:
\begin{align}
\frac{\calF (\tau_1, \tau_2, \tau_3,\tau_4)}{G(\YG{\tau_{12}})G (\YG{\tau_{34}})} = \beta\calJ  \frac{8}{\pi \alpha_K} \left[  \left( \frac{\pi \YG{\tau_{12}} }{\beta \tan  \frac{\pi \YG{\tau_{12}}}{\beta}} -1 \right) \left( \frac{\pi \YG{\tau_{34}}}{\beta \tan \frac{\pi \YG{\tau_{34}}}{\beta} } -1 \right) + \mathcal{O}(1/\beta\calJ) \right]
\end{align}
where $\alpha_K \approx 2.852$. The factorization indicates that the OPE of two fermion operators is dominated by the conserved quantity in this model --- the Hamiltonian itself, and the fluctuation of energy determine the four point function in this configuration.

Quantum chaos can be diagnosed by the out-of-time-ordered correlator (OTOC). The OTOC is calculated by starting with the Euclidean correlator with a ``j-k-j-k'' configuration of times, e.g. $\tau_1>\tau_3> \tau_2 > \tau_4$, and then giving the times $\tau_1$ and $\tau_2$ large real-time parts. Making the particular choice that all four points are evenly spaced on the imaginary time circle, we have \cite{maldacena2016comments}
\begin{align}
\calF (\frac{3\beta}{4}+ it, \frac{\beta}{4}+ it, \frac{\beta}{2},0) = \beta\calJ \frac{8}{\pi \alpha_K} \left[ (1- \frac{\pi}{2} \cosh \frac{2\pi}{\beta} t) + \mathcal{O}(1/\beta\calJ) \right] 
\label{eqn: solution for otoc}
\end{align}
which exhibits an exponential growth for real time $t$ greater than the dissipation time $t_{d}\sim \beta$, i.e., 
$\calF \sim - \beta\calJ  \exp \left( \frac{2\pi}{\beta} t \right) $ for $t\gg \beta$. This growth was recognized as a signature of chaos\cite{larkin1969quasiclassical,shenker2013black,kitaev2014hidden}, moreover, the corresponding growth exponent $\lambda_L = \frac{2\pi}{\beta}$ saturates the chaos bound proposed in Ref. \cite{maldacena2015bound}. 

The two-point function and four-point function shown above can be calculated via standard Feynman diagrams with large $N$ simplification. Another approach is to use the disorder-averaged effective action. After introducing a pair of auxiliary bi-local fields ``Green's function'' $G(\tau_1,\tau_2)$ and ``self-energy'' $\Sigma(\tau_1,\tau_2)$ and carrying out the disorder average of the partition function, the fermions can be integrated out, leaving \cite{georges2001quantum,
kitaev2015simple}:
\begin{align}
\bar{Z} &=\int \calD G  \calD \Sigma \exp\left( - N S_{\eff} [G,\Sigma  ] \right) \nn \\ 
S_{\eff}[G,\Sigma] &= -\log \Pf \left(\partial_\tau -\Sigma  \right)  + \frac{1}{2} \int d\tau_1 d\tau_2 \left( \Sigma(\tau_1,\tau_2) G(\tau_1,\tau_2)  -\frac{\calJ^2}{4}   G(\tau_1,\tau_2)^4  \right).
\end{align}
The large $N$ prefactor implies that the problem is essentially classical. The saddle point reproduces the Schwinger-Dyson equations that can also be derived via Feynman diagrams:
\begin{eqnarray}
G(i\omega) = \frac{1}{-i\omega  - \Sigma(i\omega)},\quad
\Sigma(\tau) =\calJ^2 G(\tau)^3
\label{eqn: SYK dyson equation}
\end{eqnarray}
where we assumed time translation symmetry to simplify the equations. 

One can also obtain the connected part of the four point function by considering quantum fluctuations around the saddle point. Among all the quantum fluctuations, there is a special class induced by the reparametrization of the time circle $f\in \Diff(S^1)$, which contributes the leading piece at strong coupling $\beta\calJ \gg 1$. The effective description of this part turns out to be well-captured by a local action proportional to the Schwarzian derivative\cite{kitaev2015simple,maldacena2016comments}. Remarkably, the same form of effective action also appears in the AdS$_2$ Einstein-dilaton theory\cite{Jackiw:1984je,Teitelboim:1983ux,almheiri2014models,jensen2016chaos,maldacena2016conformal,engelsoy2016investigation}.

One can also derive the thermodynamic properties from the large-$N$ saddle point free energy:
\begin{align}
\frac{F}{N}&=  \frac{1}{\beta} \left[
-\log \Pf \left(\partial_\tau -\Sigma  \right)  + \frac{1}{2} \int d\tau_1 d\tau_2 \left( \Sigma(\tau_1,\tau_2) G(\tau_1,\tau_2)  -\frac{\calJ^2}{4}   G(\tau_1,\tau_2)^4  \right) \right] \label{eqn: Free energy SYK}\\
&= U- S_0  T - \frac{\gamma }{2} T^2  + \ldots
\end{align}
In the second line we write the free energy in a low temperature expansion,\footnote{\DS{Starting at $T^{3.77}$, this expansion is expected to also involve non-integer powers given by the dimensions of irrelevant operators in the model.}} where $U\approx -0.0406 J$ is the ground state energy, $S_0 \approx 0.232$ is the zero temperature entropy\cite{georges2001quantum, kitaev2015simple}, and $\gamma T = c_v  =  \frac{\pi \alpha_K}{16\sqrt{2}\beta\calJ} \approx \frac{0.396}{\beta\calJ}$ is the specific heat\cite{maldacena2016comments}. The entropy term can be derived by inserting the conformal saddle point solution (\ref{conformalSaddle}) in the effective action. The specific heat can be derived from knowledge of the leading (in $1/\beta\calJ$) correction to the conformal saddle, but the energy requires the exact (numerical) finite $\beta \Jeff$ solution.

\section{The generalized SYK model }
\label{section: syk chain model}

In this section, we will present a simple way to generalize the SYK model to higher dimensions while keeping the solvable properties of the model in the large-$N$ limit. For concreteness of the presentation, in this section we focus on a $(1+1)$-dimensional example, which describes a one-dimensional array of SYK models with coupling between neighboring sites. It should be clear how to generalize, and we will discuss \XLQ{more details of the generalization to arbitrary dimensions and generic graphs} in section \ref{section: general models at higher dimensions}.

\subsection{Definition of the chain model}
\label{subsection: the model}

\begin{figure}
[h]
\center
\begin{tikzpicture}[scale=1.6,baseline={(current bounding box.center)}]
\draw[xshift=0pt,   fill=white] (-20pt,-33pt) rectangle (20pt,7pt);
\filldraw[xshift=0pt] (-12pt,-14pt) circle (0.6pt);
\filldraw[xshift=0pt] (-10pt,0pt) circle (.6pt);
\filldraw[xshift=0pt] (-15pt,-24pt) circle (0.6pt);
\filldraw[xshift=0pt] (15pt,-24pt) circle (0.6pt) ;
\filldraw[xshift=0pt] (-11pt,-28pt) circle (0.6pt);
\filldraw[xshift=0pt] (2pt,-22pt) circle (0.6pt);
\filldraw[xshift=0pt] (-4pt,-16pt) circle (0.6pt);
\filldraw[xshift=0pt] (-3pt,-6pt) circle (0.6pt);
\filldraw[xshift=0pt] (5pt,-8pt) circle (.6pt) ;

\filldraw[xshift=0pt] (12pt,-14pt) circle (0.6pt)
node[left] {\scriptsize $k$};
\filldraw[xshift=0pt] (12pt,-2pt) circle (.6pt) node[left]{\scriptsize $j$};

\node at (30pt,17pt) {$J'_{jklm}$};
\draw [dashed] (12pt,-14pt)--(30pt,12pt);
\draw [dashed] (12pt,-2pt)--(30pt,12pt);
\draw [dashed] (50pt,0pt)--(30pt,12pt);
\draw [dashed] (48pt,-14pt)--(30pt,12pt);

\draw[xshift=60pt, ] (-20pt,-33pt) rectangle (20pt,7pt);
\filldraw[xshift=60pt] (-3pt,-6pt) circle (0.6pt);
\filldraw[xshift=60pt] (5pt,-8pt) circle (.6pt);
\filldraw[xshift=60pt] (12pt,-2pt) circle (.6pt);
\filldraw[xshift=60pt] (12pt,-14pt) circle (0.6pt);
\filldraw[xshift=60pt] (-15pt,-24pt) circle (0.6pt);
\filldraw[xshift=60pt] (15pt,-24pt) circle (0.6pt);
\filldraw[xshift=60pt] (-11pt,-28pt) circle (0.6pt);
\filldraw[xshift=60pt] (2pt,-22pt) circle (0.6pt);
\filldraw[xshift=60pt] (-4pt,-16pt) circle (0.6pt);
\filldraw[xshift=60pt] (-12pt,-14pt) circle (0.6pt) node[right]{\scriptsize $m$};
\filldraw[xshift=60pt] (-10pt,0pt) circle (.6pt) node[right]{\scriptsize $l$};
\draw[xshift=-60pt,   fill=white] (-20pt,-33pt) rectangle (20pt,7pt);
\filldraw[xshift=-60pt] (-3pt,-6pt) circle (0.6pt) node[left]{\scriptsize $k$};
\filldraw[xshift=-60pt] (5pt,-8pt) circle (.6pt) node[right]{\scriptsize $l$};
\filldraw[xshift=-60pt] (-10pt,0pt) circle (.6pt) node[left]{\scriptsize $j$} ;
\filldraw[xshift=-60pt] (12pt,-2pt) circle (.6pt) node[right]{\scriptsize $m$};
\draw [xshift=-60pt,dashed] (12pt,-2pt)--(0pt,12pt);
\draw [xshift=-60pt,dashed] (-10pt,0pt)--(0pt,12pt);
\draw [xshift=-60pt,dashed] (5pt,-8pt)--(0pt,12pt);
\draw [xshift=-60pt,dashed] (-3pt,-6pt)--(0pt,12pt);
\filldraw[xshift=-60pt] (-12pt,-14pt) circle (0.6pt);
\filldraw[xshift=-60pt] (12pt,-14pt) circle (0.6pt);
\filldraw[xshift=-60pt] (-15pt,-24pt) circle (0.6pt);
\filldraw[xshift=-60pt] (15pt,-24pt) circle (0.6pt);
\filldraw[xshift=-60pt] (-11pt,-28pt) circle (0.6pt);
\filldraw[xshift=-60pt] (2pt,-22pt) circle (0.6pt);
\filldraw[xshift=-60pt] (-4pt,-16pt) circle (0.6pt);
\node[xshift=-90pt] at (0pt,17pt){ $J_{jklm}$};
\draw (-100pt,-13pt)--(-80pt,-13pt);
\draw (100pt,-13pt)--(80pt,-13pt);
\draw[dashed] (-100pt,-13pt)--(-120pt,-13pt);
\draw[dashed] (100pt,-13pt)--(120pt,-13pt);
\draw (-20pt,-13pt)--(-40pt,-13pt);
\draw (20pt,-13pt)--(40pt,-13pt);
\end{tikzpicture}
\caption{A chain of coupled SYK sites: each site contains $N\gg 1$ fermion with SYK interaction. The coupling between nearest neighbor sites are four fermion interaction with two from each site.}
\label{fig: chain SYK}
\end{figure}
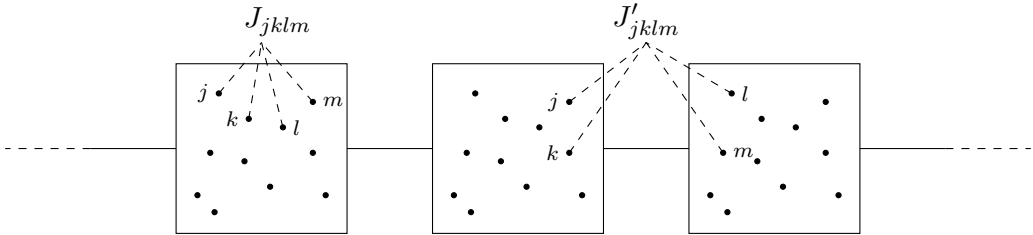

In $(1+1)$-d, our model describes a coupled array of SYK model sites, as shown in figure~\ref{fig: chain SYK}. Each site containins $N\gg 1$ Majorana fermions with SYK interactions drawn independently for each site. Each pair of neighboring sites are then further coupled via random four fermion interactions with two of the fermions from each site. The Hamiltonian has the following form:
\begin{align}
H=\sum_{x=1}^M \left( 
\sum_{1\leq j<k<l<m\leq N} J_{jklm,x} \chi_{j,x} \chi_{k,x} \chi_{l,x} \chi_{m,x} +
\sum_{\substack{1\leq j<k\leq N \\ 1\leq l<m\leq N}} J'_{jklm,x} {\chi}_{j,x} {\chi}_{k,x} {\chi}_{l,x+1} {\chi}_{m,x+1} 
\right) 
\label{eqn: Hamiltonian}
\end{align}
\YG{
where $x$ labels the lattice sites, and $\{\chi_{j,x}\}_{j=1,2, \ldots, N;~ x=1,2, \ldots, M}$ are the Majorana fermion operators satisfying anti-commutation relations and periodic boundary condition: $\{ {\chi}_{j,x},{\chi}_{k,y} \} = \delta_{xy}\delta_{jk}$, $\chi_{j,0} \equiv \chi_{j,M} $.
$N$ is the number of Majorana fermions on each site and $M$ is the number of the sites, or equivalently, the length of the chain.
In this expression, we restrict the range of indices in the sum such that each term only appears once. The first term describes the on-site SYK interaction, while the second term is the nearest neighbor random four fermion coupling.} 
The random couplings $\{J_{jklm,x}\}
$ and $\{J'_{jklm,x}\}
$ are drawn independently for each value of $x$, from a distribution with
 zero mean and variance defined in the following way:
\begin{eqnarray}
\bar{J_{jklm,x}}=\bar{J'_{jklm,x}}=0,\quad \frac{1}{3!} N^3 \bar{J^2_{jklm,x}} = \calJ_0^2,
\quad  N^3 \bar{J'^2_{jklm,x}} = \calJ_1^2.
\end{eqnarray}
The normalization, especially the factors of $N$, are chosen to make the large $N$ limit uniform and ensures the dimension 1 coupling constants $\calJ_{0}$ and $
\calJ_1$ represent the average strength of the thermal bath seen by each fermion field.

Comparing to the original SYK model, our model clearly has spatial locality. One can view our model as either $M$ coupled SYK sites, or equivalently as a big SYK site with $NM$ Majorana fermions but with inhomogeneous coupling strength---the non-local couplings (between sites $|x-y|>1$) are suppressed. As will be discussed in section~\ref{section: general models at higher dimensions}, it is not essential to have the 2-2 coupling between neighboring sites. Introducing more generic couplings such as $\chi_{j,x}\chi_{k,x+1}\chi_{l,x+2}\chi_{m,x+3}$ is straightforward as long as the coefficients of these terms are all independent variables. However, in this section we will focus on the 2-2 coupling case for simplicity. 

\begin{figure}
[h]
\center
\begin{tikzpicture}[baseline={(current bounding box.center)}]
\draw (-60pt,0pt)--(60pt,0pt);
\draw (-24pt,0pt)..controls (-10pt,15pt) and (10pt,15pt)..(24pt,0pt);
\draw (-24pt,0pt)..controls (-10pt,-15pt) and (10pt,-15pt)..(24pt,0pt);
\draw[dashed] (-24pt,0pt)..controls (-20pt,30pt) and (20pt,30pt)..(24pt,0pt);
\filldraw[fill=black](-60pt,0pt) circle (1pt) node[left]{$j,x$};
\filldraw[fill=black](60pt,0pt) circle (1pt) node[right]{$j,x$};
\filldraw[fill=black](-24pt,0pt) circle (1pt);
\filldraw[fill=black](24pt,0pt) circle (1pt);
\end{tikzpicture}
\caption{The leading order diagrams only connect fermions with same flavor and spatial coordinate under random average of disorder fields (dashed line).}
\label{fig: diagonal diagrams}
\end{figure}
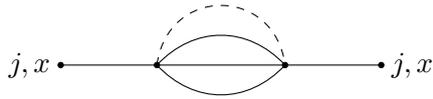

A first observation of the model is that the Hamiltonian doesn't contain any quadratic term, so that the free propagator is diagonal not only in flavor indices but also in spatial coordinate
\YG{$
\langle \calT_\tau \chi_{j,x}(\tau_1) \chi_{k,y}(\tau_2)\rangle_{\rm free} =\frac{1}{2}\delta_{jk}  \delta_{xy} \sgn  (\tau_{12}) 
$.}
Furthermore, \YG{at leading order} the interaction vertices under random average only contribute to the diagonal part, see figure~\ref{fig: diagonal diagrams}. Therefore, the dressed Green's function $\langle \calT_{\tau} \chi_{j,x}(\tau_1) \chi_{k,y}(\tau_2) \rangle$ is also diagonal in both flavor and spatial coordinates
\footnote{For the specific 2-2 interaction we choose in the chain model, the two-point functions connecting different sites vanish even without averaging over disorder, because of a $\ZZ_2$ fermion parity conservation on each site. However, even in the more general models that we will discuss in section \ref{section: general models at higher dimensions}, the cross-site two-point functions still vanish after averaging over disorder, as a consequence of local $\SO(N)$ symmetry on each site after random average.
}. In addition, the diagonal Green's functions $\langle \calT_{\tau} \chi_{j,x}(\tau_1) \chi_{j,x}(\tau_2) \rangle$ are independent of $j$ due to an $\SO(N)$ symmetry of the model after averaging over disorder. Therefore we have \YG{$\langle \calT_{\tau} \chi_{j,x}(\tau_1) \chi_{k,y}(\tau_2) \rangle=\delta_{jk} \delta_{xy} \frac{1}{N} \sum_{l=1}^N \langle \calT_{\tau} \chi_{lx}(\tau_1) \chi_{lx} (\tau_2) \rangle$. }

\subsection{The effective action and the saddle point}
\label{subsection: the effective action}
\begin{figure}
[h]
\center
\subfloat[Replicon diagonal
$\sim N$ ]
{
\begin{tikzpicture}[scale=1.4,baseline={(current bounding box.center)}]
\draw(-30pt,20pt)--(30pt,20pt);
\filldraw[black](-30pt,20pt) circle (1pt) node [left]{$\alpha$};
\filldraw[black](30pt,20pt) circle (1pt) node [right]{$\alpha$};

\draw(-30pt,20pt)..controls (-10pt,40pt) and (10pt,40pt) ..(30pt,20pt);
\draw[dashed](-30pt,20pt)..controls (-10pt,0pt) and (10pt,0pt) ..(30pt,20pt);
\draw(-30pt,20pt)..controls (-10pt,30pt) and (10pt,30pt) ..(30pt,20pt);
\draw(-30pt,20pt)..controls (-10pt,10pt) and (10pt,10pt) ..(30pt,20pt);
\end{tikzpicture}}
\hspace{40pt}
\subfloat[Off-diagonal $\sim 1/N^2$]{
\begin{tikzpicture}[scale=1.2,baseline={(current bounding box.center)}]
\draw[white] (-50pt,0pt)--(50pt,0pt) ;
\draw(-30pt,20pt)--(30pt,20pt);
\draw(-30pt,-20pt)--(30pt,-20pt);
\filldraw[black](-30pt,-20pt) circle (1pt) node [left]{$\beta$};
\filldraw[black](30pt,-20pt) circle (1pt) node [right]{$\beta$};
\filldraw[black](-30pt,20pt) circle (1pt) node [left]{$\alpha$};
\filldraw[black](30pt,20pt) circle (1pt) node [right]{$\alpha$};

\draw(-30pt,20pt)..controls (-10pt,40pt) and (10pt,40pt) ..(30pt,20pt);

\draw(-30pt,-20pt)..controls (-10pt,-40pt) and (10pt,-40pt) ..(30pt,-20pt);

\draw(-30pt,20pt)..controls (-10pt,30pt) and (10pt,30pt) ..(30pt,20pt);
\draw(-30pt,20pt)..controls (-10pt,10pt) and (10pt,10pt) ..(30pt,20pt);
\draw(-30pt,-20pt)..controls (-10pt,-30pt) and (10pt,-30pt) ..(30pt,-20pt);

\draw(-30pt,-20pt)..controls (-10pt,-10pt) and (10pt,-10pt) ..(30pt,-20pt);
\draw[dashed](-30pt,20pt)--(-30pt,-20pt);
\draw[dashed](30pt,20pt)--(30pt,-20pt);
\end{tikzpicture}}
\caption{ Replicon diagonal v.s. off-diagonal contributions to the partition function $\bar{Z^n}$: solid lines connects same replica index. Different replica indices can be connected only by dashed lines, which are disorder fields.
The replicon off-diagonal diagram on the right is suppressed by $1/N^3$ compared to the diagonal diagram on the left.
}
\label{fig: replicon diagonal diagrams}
\end{figure}
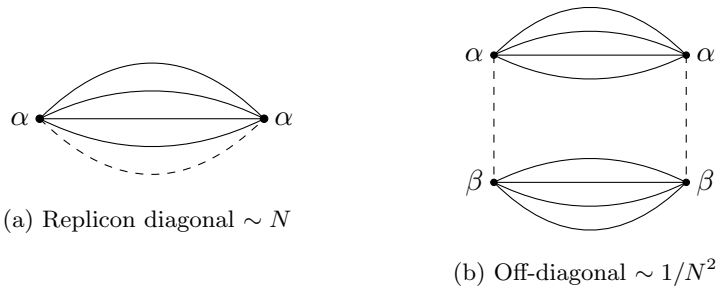

In this section, we employ the large $N$ effective action approach to analyze the model. In principle, to study the quenched problem, one should introduce $n$ replicas and analyze the disorder averaged partition function $\bar{Z^n}$. However, as in the original SYK model, our chain model self-averages at large $N$. One can verify this by checking the replicon off-diagonal contribution to the averaged partition function $\bar{Z^n}$. Figure~(\ref{fig: replicon diagonal diagrams}) shows the leading replicon off-diagonal diagram, which is suppressed by $1/N^3$ relative to the connected diagonal diagrams.  We will be interested in correlators at order $1/N$, so we can assume $\bar{Z^n}=\bar{Z}^n$ and directly work with the disorder averaged partition function and correlators. \DS{In other words, quenched equals annealed at the order we work, so we can study annealed quantities.}

To derive the effective action, we begin by integrating the partition function $Z[\{J_{jklm},J'_{jklm}\}]$ over disorder $\{ J_{jklm} \}$ and $\{ J'_{jklm} \}$ with Gaussian distributions. This will produce eight-fermion interaction terms that are non-local in time:
\begin{align}
\bar{Z}& = \int \calD \chi \exp(-S[\chi]) \nn \\
S[\chi]&= \sum_{x=1}^M  \left\lbrace \int_0^\beta  d\tau \sum_{j=1}^N \frac{1}{2} \chi_{j,x}(\tau) \partial_\tau \chi_{j,x}(\tau) 
-  \frac{1}{8N^3}  \int_0^\beta d\tau_1 d\tau_2 \left[ J_0^2 \left(\sum_{j=1}^N \chi_{j,x}(\tau_1) \chi_{j,x}(\tau_2) \right)^4 \right.\right.  \nn \\
& \left. \left.  + J_1^2 \left( \sum_{j=1}^N\chi_{j,x}(\tau_1) \chi_{j,x}(\tau_2)\right)^2 \left(\sum_{j=1}^N \chi_{j,x+1}(\tau_1) \chi_{j,x+1}(\tau_2)\right)^2  \right] \right\rbrace.
\end{align}
The expression of the averaged partition function can be further simplified by introducing bilocal auxiliary fields $G_x(\tau_1,\tau_2)$ and $\Sigma_x(\tau_1,\tau_2)$. $\Sigma_x(\tau_1,\tau_2)$ are Lagrange multipliers which impose the constraints
\begin{equation}\label{Gequation}
G_x(\tau_1,\tau_2)=\frac{1}{N} \sum_{j=1}^N \chi_{j,x} (\tau_1) \chi_{j,x} (\tau_2).
\end{equation} Integrating out the fermion fields $\chi_{jx}(\tau)$ one obtains the effective action of the bilocal fields:
\begin{align}
&~~~~~~~~~~~~~\bar{Z} = \int \calD G \calD \Sigma \exp(-N S_\eff[G,\Sigma]) \nn \\
S_{\eff}[G,\Sigma] = \sum_{x=1}^M &\left[ -\log \Pf \left(\partial_\tau -\Sigma_x  \right)  + \frac{1}{2} \int_0^\beta d\tau_1 d\tau_2 \left( \Sigma_x(\tau_1,\tau_2) G_x(\tau_1,\tau_2)   -\frac{\calJ_{0}^2}{4}   G_x(\tau_1,\tau_2)^4 \right.\right. \nn\\
&~~~~~~~~~~~~~~~~~~~~~~~~
~~~~~~~~~~~~~~~~~
~~~~\left.  \left.  - \frac{\calJ_1^2}{4} G_x(\tau_1,\tau_2)^2 G_{x+1}(\tau_1,\tau_2)^2 \right)   \right].\label{eqn: Seffchain}
\end{align}
All terms except the last term $\propto \calJ_1^2$ describe decoupled SYK models at each site, and the $\calJ_1^2$ term couples the bilocal fields $G_x(\tau_1,\tau_2)$ on neighboring sites. In the large $N$ limit, the saddle point of this action determines the leading order behavior of the two-point function $G_x(\tau_1,\tau_2)$. The saddle point equation  $\frac{\delta S_\eff}{\delta G}=0$ and $\frac{\delta S_\eff}{\delta \Sigma}=0$ produce the Schwinger-Dyson equations (assuming time translation symmetry):
\begin{align}
G_x(i\omega)&= \frac{1}{-i\omega-\Sigma_x(i\omega)}\\
\Sigma_x(\tau)=  G_x(\tau) &\left[ \calJ_{0}^2 G_{x}(\tau)^2+\frac{\calJ_1^2}{2} \left( G_{x-1}(\tau)^2   +G_{x+1}(\tau)^2 \right) \right]
\label{eqn: 2-point schwinger-dyson}
\end{align}
with $G_x(\tau)=G_x(\tau+\tau_1,\tau_1)$ and $G(i\omega)$ its fourier transformation, and similarly for $\Sigma_x(\tau)$ and $\Sigma_x(i\omega)$. This set of equations can be equivalently derived via Feynman diagrams:
\begin{eqnarray}
\begin{tikzpicture}[baseline={([yshift=2 pt]current bounding box.center)}]
\draw[thick] (-30pt,0pt)--(30pt,0pt);
\node at (0pt,-6pt){$x$};
\end{tikzpicture}
&=&\begin{tikzpicture}[baseline={([yshift= 2pt]current bounding box.center)}]
\draw (-30pt,0pt)--(30pt,0pt);
\node at (0pt,-6pt){$x$};
\end{tikzpicture}
+
\begin{tikzpicture}[baseline={([yshift=-2pt]current bounding box.center)}]
\draw (-30pt,0pt)--(10pt,0pt);
\node at (-10pt,-6pt){$x$};
\node at (20pt,-6pt){$x$};
\draw[thick] (10pt,0pt)--(30pt,0pt);
\filldraw[fill=gray] (5pt,0pt) circle (8pt);
\end{tikzpicture} 
\\
\begin{tikzpicture}[baseline={([yshift=-4pt]current bounding box.center)}]
\filldraw[fill=gray] (0pt,0pt) circle (8pt);
\node at (-15pt,0pt) {$x$};
\node at (15pt,0pt) {$x$};
\end{tikzpicture}&=& \calJ_{0}^2
\begin{tikzpicture}[scale=2,baseline={([yshift=-4pt]current bounding box.center)}]
\draw [thick] (-12pt,0pt)--(12pt,0pt);
\draw [thick](-12pt,0pt)..controls (-4pt,8pt) and (4pt,8pt)..(12pt,0pt);
\draw [thick](-12pt,0pt)..controls (-4pt,-8pt) and (4pt,-8pt)..(12pt,0pt);
\node at (0pt,8pt) {\scriptsize $x$};
\node at (0pt,2pt) {\scriptsize $x$};
\node at (0pt,-9pt) {\scriptsize $x$};
\end{tikzpicture}  + \frac{\calJ_1^2}{2} \left(
\begin{tikzpicture}[scale=2,baseline={([yshift=-4pt]current bounding box.center)}]
\draw [thick] (-12pt,0pt)--(12pt,0pt);
\draw [thick](-12pt,0pt)..controls (-4pt,8pt) and (4pt,8pt)..(12pt,0pt);
\draw [thick](-12pt,0pt)..controls (-4pt,-8pt) and (4pt,-8pt)..(12pt,0pt);
\node at (0pt,8pt) {\scriptsize $x-1$};
\node at (0pt,2pt) {\scriptsize $x-1$};
\node at (0pt,-9pt) {\scriptsize $x$};
\end{tikzpicture} + 
\begin{tikzpicture}[scale=2,baseline={([yshift=-4pt]current bounding box.center)}]
\draw [thick] (-12pt,0pt)--(12pt,0pt);
\draw [thick](-12pt,0pt)..controls (-4pt,8pt) and (4pt,8pt)..(12pt,0pt);
\draw [thick](-12pt,0pt)..controls (-4pt,-8pt) and (4pt,-8pt)..(12pt,0pt);
\node at (0pt,8pt) {\scriptsize $x+1$};
\node at (0pt,2pt) {\scriptsize $x+1$};
\node at (0pt,-9pt) {\scriptsize $x$};
\end{tikzpicture}  \right),
\label{eqn: diagram self energy}
\end{eqnarray}
where the thick lines represent dressed Green's functions and the gray disk represents the self-energy.

Notice that the chain model under disorder average is translation invariant and also has a translation invariant free propagator $G^{\rm free}_x(\tau)=\frac{1}{2}\sgn(\tau)$. Therefore we consider translation invariant solutions of the Schwinger-Dyson equations:
\begin{align}
G_x(\tau)=G^s(\tau),&\quad \Sigma_x(\tau)=\Sigma^s(\tau); \\
G^s(i\omega)= \frac{1}{-i\omega-\Sigma^s(i\omega)},&\quad \Sigma^s(\tau)= (  \calJ_{0}^2+\calJ_1^2) G^s(\tau)^3.  \label{eqn: chain Dyson}
\end{align}
Comparing to equation~(\ref{eqn: SYK dyson equation}), one sees the Schwinger-Dyson equations reduce to exactly the same form as those of a $(0+1)$-d SYK model 
with the coupling constant $\calJ= \sqrt{\calJ_{0}^2+\calJ_1^2}$. Therefore, we can directly apply the SYK model results in Ref.~\cite{kitaev2015simple,maldacena2016comments}. In particular, we immediately know that the solution in the conformal limit $N \gg \beta \calJ \gg 1$ has the following familiar form:
\begin{align}
G^s\left( \tau_1,\tau_2 \right)
=&~ b^{\Delta}  \left(  \frac{\beta\Jeff }{\pi} \sin \frac{\pi \YG{\tau_{12}}}{\beta} \right)^{-2\Delta},\quad 0\leq \YG{\tau_{12}} < \beta  \label{eqn: conformal two point function}
\\
  b=&~ \frac{1}{\pi } \left(\frac{1}{2}-\Delta \right) \tan (\pi \Delta),\quad  \Delta=1/4 \nn.
\end{align}
Here and below, we will denote the effective coupling $\sqrt{\calJ_{0}^2+\calJ_1^2}$ in our model as $\Jeff$.

In summary, the two-point function in our chain model is local in space and power-law decaying (at low temperature) in time, a behavior known as local quantum criticality\cite{si2001locally,faulkner2011emergent}. This saddle point solution (and the finite $\beta J$ corrections to it) is the starting point for studying other properties of the model.

For example, inserting the solution into the action, we derive the saddle point approximation to the partition function. This gives the order $N$ term in the free energy:
\begin{align}
\frac{F}{N M} &= \frac{1}{\beta} \left[ -\log \Pf \left(\partial_\tau -\Sigma^s  \right)  + \frac{1}{2} \int d\tau_1 d\tau_2 \left( \Sigma^s(\tau_1,\tau_2) G^s(\tau_1,\tau_2)  -\frac{\Jeff^2  }{4}   G^s(\tau_1,\tau_2)^4  \right)\right].
\end{align}
Here we have used space translation symmetry to simplify the notation. ($M$ is the number of lattice sites, { i.e.} the spatial volume of the system.) Using the saddle point solution one can see that the free energy density $\frac{F}{NM}$ agrees exactly with that in the SYK model Eq.~(\ref{eqn: Free energy SYK}) with $\calJ=\sqrt{\calJ_0^2+\calJ_1^2}$. 
Therefore, in the large $N$ limit our model has exactly the same zero temperature entropy per fermion $S_0\approx 0.232$, and specific heat $c_v \approx \frac{0.396}{\beta\calJ}$ per fermion. It should be noted that the zero temperature entropy is extensive. 

We should remark here that the exact agreement on thermodynamic properties with the SYK model only holds at leading order in the $1/N$ expansion. This is similar and related to the discussion of the two-point function, where the agreement with the SYK model only holds at leading order. We will see below that $1/N$ effects, which are the quantum fluctuations around the large $N$ saddle point, are different in our chain model than in the SYK model.

\subsection{Fluctuations of the collective fields and four-point functions}

The fermion four-point function can be determined from the fluctuations about the saddle point just discussed. At large $N$ the saddle is sharp, and the connected four point function is small, of order $\frac{1}{N}$. This is determined by the Gaussian fluctuations of bilocal fields $G_x(\tau_1,\tau_2)$ and $\Sigma_x(\tau_1,\tau_2)$ about the saddle. 

It is convenient to expand about the saddle using variables $g_x,\sigma_x$ defined by
\begin{align}
G_x(\tau_1,\tau_2)&=G^s(\tau_1,\tau_2)+ |G^s(\tau_1,\tau_2)|^{-1}  g_x(\tau_1,\tau_2),\nn \\ 
\Sigma_x(\tau_1,\tau_2)&=\Sigma^s(\tau_1,\tau_2) + |G^s(\tau_1,\tau_2)| \sigma_x(\tau_1,\tau_2),
\end{align}
where we have rescaled the fluctuation fields $g_x,\sigma_x$ by prefactors ${|G^s|}^{-1}$ and ${|G^s|}$ for convenience. It should be noticed that although the saddle point is uniform in space and translation invariant in time, the fluctuation fields have generic space-time dependence. Using Eq.~(\ref{Gequation}), the connected, averaged four-point function of the fermions (defined analogously to Eq.~(\ref{fourPtDef})) can be written as the connected two-point function of $g_x(\tau_1,\tau_2)$:
\begin{eqnarray}
\frac{1}{N}\calF_{xy} (\tau_1,\tau_2; \tau_3,\tau_4) &=& \langle G_x(\tau_1,\tau_2) G_y(\tau_3,\tau_4) \rangle - \langle G_x(\tau_1,\tau_2) \rangle \langle G_y(\tau_3,\tau_4) \rangle \nonumber\\
&=&|G^s(\tau_{12})|^{-1}|G^s(\tau_{34})|^{-1}\langle g_x(\tau_1,\tau_2)g_y(\tau_3,\tau_4)\rangle.
\end{eqnarray}
More precisely, $\frac{\mathcal{F}}{N}$ is the connected part of the fermion four point function. We will see that $\mathcal{F}$ is of order one at large $N$, so the connected correlator is of order $\frac{1}{N}$.

To compute the $\langle g_x g_y\rangle$ correlator, we expand the effective action to second order in the fluctuation fields $g, \sigma$, which leads to
\begin{align}
\delta S_{\eff}[g,\sigma] = & -\frac14  \int d^4\tau \sum_{x} \sigma_x(\tau_1,\tau_2)G^s(\tau_{13})\cdot |G^s(\tau_{34})|\cdot G^s(\tau_{42})\cdot |G^s(\tau_{21})|\sigma_x(\tau_3,\tau_4)\nonumber\\
& +\int d^2\tau \left(\sum_x\frac12\sigma_x(\tau_1,\tau_2)g_x(\tau_1,\tau_2) -\frac{3\Jeff^2}{4}\sum_{x,y}g_x(\tau_1,\tau_2)S_{xy}  g_y(\tau_1,\tau_2) \right).
\end{align}
The spatial kernel $S_{xy}$ is a tight-binding hopping matrix
\begin{eqnarray}
S_{xy}&=& \delta_{x,y} +\frac{\calJ_1^2 }{3\Jeff^2}\left(\delta_{x,y\pm1} -2\delta_{x,y}\right)\label{eqn: spatial kernel},
\end{eqnarray}
where we continue to use the notation that $\Jeff=\sqrt{\calJ_0^2+\calJ_1^2}$.

It is straightforward to integrate out $\sigma_x$ and obtain a quadratic action for $g_x$ alone. We define 
$\tilde{K}$ as the (symmetrized) four-point function kernel of the SYK model\cite{kitaev2015simple,maldacena2016comments}:
\begin{eqnarray}
\tilde{K}\left(\tau_1,\tau_2;\tau_3,\tau_4\right)= 3\Jeff^2 G^s(\tau_{13}) \cdot|G^s(\tau_{34})| \cdot G^s(\tau_{42})\cdot|G^s(\tau_{21})|.
\end{eqnarray}
The effective action of $g_x$ is
\begin{align}
\delta S_{\eff}[g] =\frac{3\Jeff^2}{4}\int d^4\tau\sum_{x,y}g_x(\tau_1,\tau_2)\left[\tilde{K}^{-1}\left(\tau_1,\tau_2;\tau_3,\tau_4\right)\delta_{xy}-S_{xy}\delta(\tau_{13})\delta(\tau_{24})\right]g_y(\tau_3,\tau_4),
\label{eqn: effective action for fluctuations}
\end{align}
which determines the fermion four-point function:
\begin{eqnarray}
\frac1N\calF_{xy}\left(\tau_1,\tau_2;\tau_3,\tau_4\right)&=&\frac1{|G^s(\tau_{12})G^s(\tau_{34})|}\langle g_x(\tau_1,\tau_2)g_y(\tau_3,\tau_4)\rangle\nonumber\\
&=&\frac1N\frac1{|G^s(\tau_{12})G^s(\tau_{34})|}\frac2{3\Jeff^2}\left(\tilde{K}^{-1}-S\right)^{-1}.
\label{eq:fourptgeneral}
\end{eqnarray}
Here we have written the correlator in a compact matrix form.

Comparing Eq.~(\ref{eq:fourptgeneral}) with the four-point function of the SYK model, the only difference is with the spatial kernel $S$. Replacing $S_{xy}$ by $\delta_{xy}$ ({i.e.} taking $\calJ_1=0$) reproduces the SYK result, as expected.

The behavior of the four-point function (\ref{eq:fourptgeneral}) can be analyzed by diagonalizing the kernel $(\tilde{K}^{-1}-S)$. First of all, due to translation symmetry it is straightforward to do a spatial Fourier transformation to $(x-y)$ and define the Fourier component 
\begin{eqnarray}
\frac1N\calF_p(\tau_1,\tau_2;\tau_3,\tau_4)&=&\frac1N\frac1{|G^s(\tau_{12})G^s(\tau_{34})|}\frac2{3\Jeff^2}\left[\tilde{K}^{-1}-s(p)\delta\left(\tau_{13}\right)\delta\left(\tau_{24}\right)\right]^{-1}\nonumber\\
\text{with~}s(p)&=&1+\frac{2\calJ_1^2}{3\Jeff^2}\left(\cos p-1\right).\label{eq:sp}
\end{eqnarray}
Then one can diagonalize the temporal kernel $\tilde{K}$ in the same way as in the SYK model. We write the antisymmetric eigenfunctions $\Psi_{h,n}\left(\tau_1,\tau_2\right)$ where $n$ labels the fourier mode for the sum of the two times, and $h$ specifies the dependence on the difference of the times. Writing the corresponding eigenvalues of $\tilde{K}$ as $k(h,n)$, the four-point function can be expressed as
\begin{align}
\frac1N\calF_p(\tau_1,\tau_2;\tau_3,\tau_4)=\frac1N\frac1{|G^s(\tau_{12})G^s(\tau_{34})|}\frac2{3\Jeff^2}\sum_{h,n}\Psi_{h,n}(\tau_1,\tau_2)\frac{k(h,n)}{1-s(p)k(h,n)}\Psi_{h,n}^*(\tau_3,\tau_4)
\label{eqn: general formula for finite $p$}
\end{align}
(where we have used the fact that the symmetrized kernel is Hermitian). The only difference from the original SYK model is the factor of $s(p) \leq 1$ in the denominator.\footnote{\DS{It is interesting to note that the same type of modification was found in a different generalization of the SYK model in appendix G of \cite{maldacena2016comments}.}} The details of the eigenvectors and eigenvalues of the temporal kernel, $k(h,n)$ and $\Psi_{h,n}$ have been worked out in Ref.~\cite{maldacena2016comments}, and we will use them in the following sections. It should be noted that Eq.~(\ref{eqn: general formula for finite $p$}) returns to the SYK result at zero momentum $p=0$, as one expects from the form of the effective action. 

\subsection{Symmetry breaking and pseudo-Goldstone mode}

Although the general discussion above is sufficient for calculating four-point functions, it is helpful to gain more physical understanding by analyzing symmetry properties of the saddle point solution.
\begin{figure}
[t]
\center
\subfloat[Green's function]
{
\begin{tikzpicture}[scale=0.6,baseline={(current bounding box.center)}]
\draw (0pt,0pt) circle (50pt);
\filldraw[fill=black]  (-30pt,40pt) circle (1pt)  node [left] {$\tau_1$};
\filldraw[fill=black] (30pt,40pt) circle (1pt)   node [right] {$\tau_2$};
\draw [dashed] (-30pt,40pt) -- (30pt,40pt);
\node at (0pt,30pt){\scriptsize $G_x(\tau_1,\tau_2)$};
\filldraw[fill=black] (50pt,0pt) circle (1pt) node [above right]{$0$} node [below right]{$\beta$} ;
\node at (0pt,-58pt) {$x$};
\end{tikzpicture}
}
\subfloat[Space-time picture for chain model]
{
\begin{tikzpicture}[scale=1.1,baseline={(current bounding box.center)}]
\draw [dashed] (-100pt,20pt) -- (80pt,20pt);
\draw  (-80pt,20pt) -- (60pt,20pt);

\draw (-55pt,0pt) ellipse (8pt and 20pt);
\draw (-25pt,0pt) ellipse (8pt and 20pt);
\draw (5pt,0pt) ellipse (8pt and 20pt);
\draw (35pt,0pt) ellipse (8pt and 20pt);

\draw[<-,>=stealth] (-32pt,25pt) ..controls (-27pt, 30pt) and (-22pt,30pt) .. (-18pt,25pt) node[right]{$\tau$};
\draw [dashed] (-100pt,-20pt) -- (80pt,-20pt);
\draw  (-80pt,-20pt) -- (60pt,-20pt);
\draw[->,>=stealth]  (-40pt,-30pt) -- (20pt,-30pt) node [right]{$x$};
\end{tikzpicture}
}

\subfloat[Analogy to ferromagnetic spin chain with pinning field]{
\begin{tikzpicture}[scale=1.8, baseline={(current bounding box.center)}]
\draw (-10pt,0pt)--(100pt,0pt);
\filldraw (0pt,0pt) circle (1pt);
\filldraw (15pt,0pt) circle (1pt);
\filldraw (30pt,0pt) circle (1pt);
\filldraw (45pt,0pt) circle (1pt);
\filldraw (60pt,0pt) circle (1pt);
\filldraw (75pt,0pt) circle (1pt);
\filldraw (90pt,0pt) circle (1pt);
\draw[rotate=15,->,>=stealth] (0pt,-10pt)--(0pt,10pt);
\draw[xshift=15pt,rotate=10,->,>=stealth] (0pt,-10pt)--(0pt,10pt);
\draw[xshift=30pt,rotate=5,->,>=stealth] (0pt,-10pt)--(0pt,10pt);
\draw[xshift=45pt,rotate=-10,->,>=stealth] (0pt,-10pt)--(0pt,10pt);
\draw[xshift=60pt,rotate=0,->,>=stealth] (0pt,-10pt)--(0pt,10pt);
\draw[xshift=75pt,rotate=-5,->,>=stealth] (0pt,-10pt)--(0pt,10pt);
\draw[xshift=90pt,rotate=5,->,>=stealth] (0pt,-10pt)--(0pt,10pt);

\draw[xshift=-25pt,rotate=0,thick, blue,->,>=stealth] (0pt,-10pt)--(0pt,10pt) node[above] {$B_z$};
\end{tikzpicture}
}
\caption{(a) Green's function $G_x(\tau_1,\tau_2)$ is a function of two imaginary time variables, each defined on the imaginary time circle. It transforms covariantly under the reparametrization field $f_x\in \Diff(S^1)$. (b) The space-time picture for the chain model. The Schwinger-Dyson equation at conformal limit has global reparametrization symmetry, but the conformal solution spontaneously breaks $\Diff(S^1)$ to $\PSL_2(\RR)$. Moreover, the UV term $-i\omega$ in (\ref{eqn: chain Dyson}) breaks the emergent reparametrization and lifts the Goldstone modes to pseudo-Goldstone modes. (c) The situation in the SYK chain model is similar to a ferromagnetic spin chain with a small pinning field, where the $\SU(2)$ symmetry is ``almost spontaneously" broken to $\UU(1)$, leading to a pseudo-Goldstone mode. 
}
\label{fig: spin chain analogy}
\end{figure}
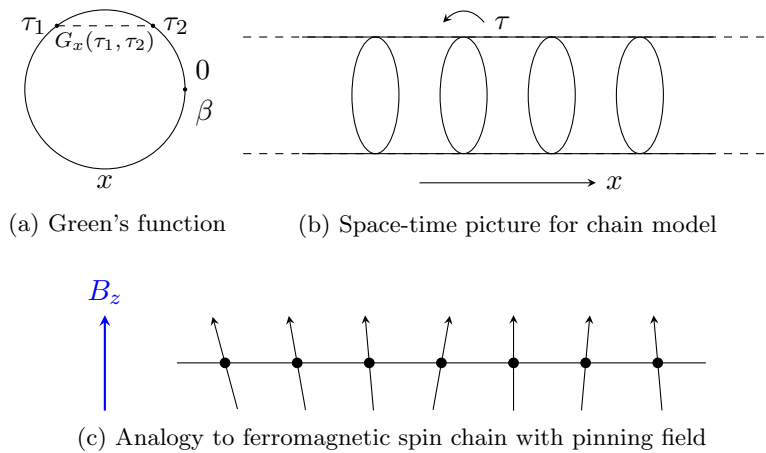
As in the SYK model, our effective action (\ref{eqn: Seffchain}) admits an approximate reparametrization symmetry of time in the IR limit $\omega\rightarrow 0$, where one can ignore the $-i\omega$ in the first term. The reparameterization transformation is defined by
\begin{equation}
f\in \Diff(S^1): \quad G_x(\tau_1,\tau_2) \rightarrow \left( f'(\tau_1) f'(\tau_2)  \right)^\Delta	G_x(f(\tau_1),f(\tau_2)).
\end{equation}
\XLQ{The symmetry $\Diff(S^1)$ is broken spontaneously to $\PSL_2(\RR)$ by the solution in equation~(\ref{eqn: conformal two point function}). If this symmetry were exact, the spontaneous symmetry breaking would have produced infinite number of Goldstone modes corresponding to the spatially dependent
 reparametrizations $f_x\in \Diff(S^1)/\PSL_2(\RR)$. The $-i\omega$ term in the effective action (\ref{eqn: chain Dyson}) plays the role of a small explicit symmetry breaking field, which selects the solution in Eq. \ref{eqn: conformal two point function} and turns the Goldstone bosons to pseudo-Goldstone bosons. This is similar to the situation in a ferromagnetic spin chain with small external magnetic field (see figure~\ref{fig: spin chain analogy}).} 
\DS{The effect of this symmetry breaking term is small at large $\beta \Jeff$, which means that the pseudo-Goldstone bosons have large fluctuations and make the most important contributions to the long-wavelength dynamics. In particular, they are} \YG{responsible for the diffusion of energy that we will analyze in section~\ref{section: transport} and the chaos characteristics we will study in section~\ref{section: chaos and the butterfly velocity}.}

From the view point of the effective action, the pseudo-Goldstone modes are those fluctuations along the ``nearly-flat'' direction around the saddle point in the potential of $S_{\eff}[G]$. As we have discussed, these fluctuations correspond to the spontaneously and explicitly broken reparametrization symmetry, and therefore can be parametrized by residue target space $\Diff(S^1)/\PSL_2(\RR)$ at each point in space. More explicitly, $f_x$ acts on the saddle point in the following way
\begin{align}
\YG{f_x(\tau)\in \Diff(S^1)} , \quad
G^s(\tau_1,\tau_2)\rightarrow G^f_x(\tau_1,\tau_2):=\left( f'_x(\tau_1) f'_x(\tau_2)  \right)^\Delta	G^s(f_x(\tau_1),f_x(\tau_2)).
\end{align}
For small deformations of time $f_x(\tau)=\tau+\epsilon_x(\tau)$, the quadratic action for $\epsilon_x(\tau)$ can be determined by diagonalizing the kernel $\tilde{K}$ and using (\ref{eqn: effective action for fluctuations}). Building on the results of \cite{maldacena2016comments}, we will find below that to quadratic order in the spatial momentum $p$, this leads to the action
\begin{align}
S = \frac{1}{256\pi}  \sum_{n,p} \epsilon_{n,p} \left( \frac{\sqrt{2}\alpha_K }{\beta \Jeff }  n^2(n^2-1) +  \frac{\calJ_1^2}{3\Jeff^2} p^2 |n|(n^2-1)   \right) \epsilon_{-n,-p}~,\label{eqn: reparametrizationaction}\\
f_x(\tau) = \tau+ \epsilon_x(\tau), \hspace{20pt} \epsilon_{n,p} = \frac{1}{\sqrt{M}}\sum_{x=1}^M \int_0^\beta d\tau e^{i(\frac{2\pi n}{\beta}\tau-xp)}\epsilon_x(\tau).
\end{align}
The first term is familiar from the SYK model. It is local in time, and can be interpreted as a quadratic approximation to the Schwarzian derivative action at each site. The coefficient $\frac1{\beta \Jeff}$ tells us that at large $\beta\Jeff$ limit the reparameterization fields are soft modes, due to the approximate reparameterization symmetry. The second term describes a simple coupling of the reparameterization modes at different sites, but with a nonlocal form as a function of time.\footnote{An interesting question is what is the full non-linear form of the effective action which generalizes the Schwarzian action in $(0+1)$-d case, which we will leave for future work.} We will see that together, these two terms determine both the energy diffusion dynamics and chaos behavior.\footnote{Notice that this action has three zero modes, $n = -1,0,1$ for each point in space. These are set to zero by the quotient in $\Diff(S^1)/\PSL_2(\RR).$}

\begin{figure}
[ht]
\center
\subfloat[j-j-k-k order]{
\begin{tikzpicture}[baseline={(current bounding box.center)}]
\draw [->,>=stealth] (-20pt,0pt)--(100pt,0pt) node[right]{\small space};
\draw [->,>=stealth] (0pt,-20pt)--(0pt,80pt) node[above]{\small imaginary time};
\filldraw (80pt,60pt) circle (1pt) node[right] {$\tau_1$};
\filldraw (80pt,50pt) circle (1pt) node[right] {$\tau_2$};
\filldraw (0pt,10pt) circle (1pt) node[above left] {$\tau_3$};
\filldraw (0pt,0pt) circle (1pt) node[above left] {$\tau_4$};
\filldraw (80pt,0pt) circle (1pt) node[below right] {$x$};
\draw[dashed] (80pt,-10pt) -- (80pt,70pt);
\end{tikzpicture}
}
\subfloat[j-k-j-k order]{
\begin{tikzpicture}[baseline={(current bounding box.center)}]
\draw [->,>=stealth] (-20pt,0pt)--(100pt,0pt) node[right]{\small space};
\draw [->,>=stealth] (0pt,-20pt)--(0pt,80pt) node[above]{\small imaginary time};
\filldraw (80pt,60pt) circle (1pt) node[right] {$\tau_1$};
\filldraw (80pt,20pt) circle (1pt) node[right] {$\tau_2$};
\filldraw (0pt,40pt) circle (1pt) node[above left] {$\tau_3$};
\filldraw (0pt,0pt) circle (1pt) node[above left] {$\tau_4$};
\filldraw (80pt,0pt) circle (1pt) node[below right] {$x$};
\draw[dashed] (80pt,-10pt) -- (80pt,70pt);
\end{tikzpicture}
}
\caption{Two regions of the four-point function are illustrated. Factorization in the configuration at left gives the propagating bilocal fields. The configuration at right can be continued to the OTOC which diagnoses chaos.}
\label{fig: different region of 4pt function}
\end{figure}
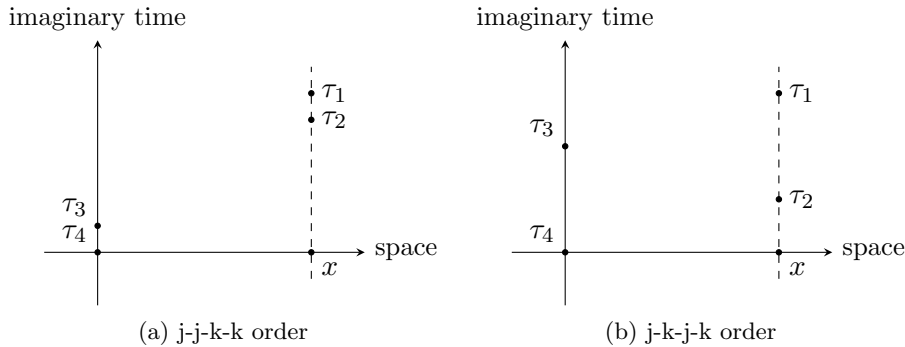

In the following two sections, we will analyze two different regions of the four-point function with different ordering of the time variables $\tau_1,\tau_2,\tau_3,\tau_4$ as shown in figure~(\ref{fig: different region of 4pt function}). The four-point functions with ordering $\tau_1>\tau_2>\tau_3>\tau_4$, which we will sometimes refer to as j-j-k-k order, determines the dynamics of collective fields in our model, while the order $\tau_1>\tau_3>\tau_2>\tau_4$ which we will refer to as j-k-j-k determines OTOC after analytic continuation, and characterizes chaos in our system.

\section{The OPE region and the Energy Transport}
\label{section: transport}

As we have seen from the two-point functions and four-point functions, single fermion fields $\chi_{jx}(\tau)$ do not propagate spatially in our model. The only fields that have nontrivial dynamics are the collective bilocal fields $g_x(\tau_1,\tau_2)$, which are singlet in the flavor $\SO(N)$ symmetry. In the four-point function $\calF_{xy}\left(\tau_1,\tau_2;\tau_3,\tau_4\right)$, if we denote $\tau=\frac12\left({\tau_1+\tau_2-\tau_3-\tau_4}\right)$, and take the limit $|\tau_2-\tau_1|,|\tau_4-\tau_3|\ll |\tau|$, we are effectively taking an operator product expansion (OPE) of the fermion fields $\chi_{j,x}(\tau_1)\chi_{j,x}(\tau_2)$ and similarly for $\chi_{k,y}(\tau_3)\chi_{k,y}(\tau_4)$. The four-point function becomes a sum over the propagators of different collective fields that appear in the OPE. Roughly speaking, different collective fields correspond to different families of eigenvalues in Eq.~(\ref{eqn: general formula for finite $p$}). 

The eigenvalues and eigenfunctions of the temporal kernel have been studied in Ref.~\cite{kitaev2015simple,maldacena2016comments}. In the strong coupling limit $\beta\Jeff \gg 1$, the two-point functions take the conformal form (\ref{eqn: conformal two point function}). This correlation function is covariant under M\"obius transformations ($\PSL_2(\RR)$) of the time coordinate, which is used in diagonalizing the temporal kernel. It turns out\cite{kitaev2015simple,
polchinski2016spectrum,
maldacena2016comments} that the eigenvalues of the kernel are given by the simple formula
\begin{equation}\label{confeigs}
k(h,n)\rightarrow k(h)= -\frac{3 }{2h-1} \tan \frac{\pi}{2} (h-\frac{1}{2}),\quad h\in \{2,4,6,\ldots, \frac{1}{2}+ i \RR^+ \} = I.
\end{equation}
The label set $I$ consists of a discrete series and a continuum\cite{
polchinski2016spectrum,
maldacena2016comments}. For small momentum $p^2 \lesssim \frac{\Jeff}{\beta \Jeff_1^2}$, we will see that $\frac{1}{\beta\Jeff}$ corrections to this formula are important. First, however, we study the case of finite $p$ where the conformal large $\beta \Jeff$ limit is straightforward.

 \subsection{The OPE at finite $p$}
 \label{subsection: conformal}

The fact that the eigenvalues (\ref{confeigs}) don't depend on $n$ is a consequence of the conformal symmetry of the eigenvalue problem at large $\beta \Jeff$. It makes it possible to do the sum over $n$ in Eq.~(\ref{eqn: general formula for finite $p$}), which gives particular hypergeometric functions of the cross ratio
\begin{equation}\label{crossRatioDef}
\eta= \frac{\sin\frac{\pi\tau_{12}}{\beta}\sin\frac{\pi\tau_{34}}{\beta}}{\sin\frac{\pi\tau_{13}}{\beta}\sin\frac{\pi\tau_{24}}{\beta}}.
\end{equation}
The sum and integral over the remaining eigenvector index $h\subset I$ can then be done following Ref.~\cite{maldacena2016comments} by deforming the contour of integration over the continuum portion of $I$. As the contour is pushed to infinity, one encounters two sets of poles: one set conveniently cancels the contribution from the discrete $h\subset I$ and the other gives the answer (in the region $\tau_1\YG{>}\tau_2\YG{>}\tau_3\YG{>}\tau_4$ where $\eta<1$)
\begin{equation}\label{fourptres}
\calF_{p}= - {G^s(\tau_{12})G^s(\tau_{34})}\frac{4\pi}{3} \sum_{m=0}^\infty \operatorname{Res} \left[ \frac{h-1/2}{\pi  \tan \pi h/2 } \frac{k(h)}{1-s(p)k(h)} \frac{\Gamma(h)^2}{\Gamma(2h)} \eta^h ~_2F_1(h,h,2h,\eta) \right]_{h=h_{m}(p)}
\end{equation}
The only difference in (\ref{fourptres}) from the regular SYK case \cite{maldacena2016comments} is that we have the factor of $s(p)$. Here$~_2F_1(h,h,2h,\eta)$ is the hypergeometric function, which is regular and approaches constant $1$ at small $\eta$. 
The values $\{h_m(p)\}$ that we sum over are the infinite set of positive solutions of the pole equation $1-k(h)s(p)=0$. They are approximately double-integer spaced, approaching $h_m(p) \rightarrow \frac{3}{2} + 2m$ for large $m$. 

Taking the residues, we have
\begin{align}
\calF_p &= -G^s(\tau_{12})G^s(\tau_{34})\sum_{m} c^2_m(p) \eta^{h}{}_2F_1(h,h,2h,\eta)\Big|_{h = h_m(p)}\\
c^2_m(p)&=\frac{4\pi}{3 k'(h)} \frac{h-1/2}{\tan \pi h/2} \frac{\Gamma(h)^2}{\Gamma(2h)} \frac{1}{s(p)^2}\Big|_{h = h_m(p)} \label{OPEcoeff}.
\end{align}
If one considers the OPE limit $\Jeff^{-1}\ll\tau_{12},\tau_{34}\ll \tau=\tau_{13}\simeq \tau_{24}$, we have $\eta 
\simeq \frac{\tau_{12}\tau_{34}}{(\frac{\beta}{\pi}\sin\frac{\pi\tau}{\beta})^2}$, and
\begin{align}
&\calF_{p} \propto G^s(\tau_{12})G^s(\tau_{34})\sum_{m} c^2_m(p) \eta^{h_m(p)}\simeq \sum_m \tau_{12}^{h_m(p)-2\Delta}\tau_{34}^{h_m(p)-2\Delta}c^2_m(p) \left(\frac{\beta}{\pi}\sin\frac{\pi\tau}{\beta}\right)^{-2h_m(p)},\label{OPEofconformal}
\end{align}
with $\Delta=\frac14$ the dimension of the fermion field. For fixed $p$, the dimension $h_m(p)$ labels the time $\PSL_2(\RR)$ representation of the operator $\phi_m$ appearing in the OPE.

It is also useful to consider the fourier transform of $\calF_p$ back to position space.\footnote{Here we consider only the modes $m\geq 1$ that have smooth behavior for $p\rightarrow 0$. We will consider the $m  = 0$ (i.e. $h \approx 2$) contribution in detail below.} One can show that $c_m^2(p)$ is analytic in a strip surrounding the real $p$ axis, so the fourier transform will decay exponentially in $x$. More precisely, $c_m^2(p)$ has both a pole and a branch cut for complex $p$. The pole is at the location where $h_m(p)$ becomes an even integer, and the $\tan(\pi h_m/2)$ in the denominator diverges. The branch cut is due to the multivaluedness of the solution to $k(h_m) = s(p)^{-1}$ and occurs at the location where $k'(h_m(p)) = 0$. For large $m$, one can show that both singularities are at a distance of order $\log m$ from the real $p$ axis, so the contribution of the $m$ mode will decay in space as $e^{-\log(m) |x|}$ for large $x$ and large $m$. Here $x$ is measured in lattice units, so the modes only propagate a few lattice sites. However, note that the decay factor in the exponential, $\log (m) |x| \sim \log(h) |x|$, grows less rapidly with $h$ than in a conformally invariant $(1+1)$ dimensional theory, where we would have $h|x|$.

\subsection{The $h=2$ contribution for small $p$}
\label{subsection: h=2}
For small momentum $p$, the first solution to $k(h_m)s(p) = 1$ approaches $h_0(0) = 2$. Because of the factor of $\tan\frac{\pi h_m}{2}$ in the denominator of Eq.~(\ref{OPEcoeff}), this leads to a large OPE coefficient $c_0^2(p)$ that diverges like $p^{-2}$ as $p\rightarrow 0$. It will be important to understand how this is cut off at finite $\beta \Jeff$.

The would-be divergence can be traced to the $h = 2 \subset I$ eigenfunctions of $\tilde{K}$. In this section we study the $h=2$ piece in more detail, which requires taking into account the non-conformal corrections to $k(h,n)$. The correction can be derived by including the $1/\beta\Jeff$ correction to the conformal form of the two-point function, as has been discussed in Ref.~\cite{maldacena2016comments}. In the following we will apply the SYK results to our generalized model.

To start, we note that due to time translation symmetry, the eigenfunctions $\Psi_{2}\left(\tau_1,\tau_2\right)$ for $h=2$ have the form $\Psi_2\left(\tau_1,\tau_2\right)=\Psi_{2,n}(\tau_{12})e^{-i\frac{2\pi}\beta n\frac{\tau_1+\tau_2}2}$. In the conformal limit $\beta\Jeff\gg1$, $\Psi_{2,n}(\tau_{12})$ has the following explicit form:
\begin{align}
\Psi_{2,n}\YG{(\tau_{12})} = \frac{\gamma_n}{2\sin \YG{ \frac{\pi \tau_{12}}{\beta} }} f_n(\YG{\tau_{12}}),\quad f_n(\YG{\tau_{12}}):=\frac{\sin n \YG{ \frac{\pi \tau_{12}}{\beta} }}{\tan \YG{ \frac{\pi \tau_{12}}{\beta} }} -n \cos n \YG{ \frac{\pi \tau_{12}}{\beta} }.
\end{align}
In the above expression, $\gamma_n^2=\frac{3}{\pi^2 |n|(n^2-1)}$ is the prefactor to normalize $\Psi_{2,n}$. In the first order perturbation theory, one can use these zeroth order (conformally covariant) eigenfunctions to obtain the first order corrections to the eigenvalues, which are given in Ref.~\cite{maldacena2016comments} by 
\begin{eqnarray}
k(h=2,n)\simeq 1-\frac{\sqrt{2}\,\alpha_K}{\beta\Jeff}|n|+...
\end{eqnarray}
where $\alpha_K \approx 2.852$. Summing over the eigenfunctions with this improved formula for the eigenvalue, we find the $h = 2$ contribution
\begin{align}
\frac{\calF_{p,h=2}(\tau,\tau_{12},\tau_{34})}{G^{s}(\tau_{12})G^{s}(\tau_{34})}
&\simeq \frac{8}{\pi}\sum_{n} \frac{e^{-i\omega_n\tau}}{1-s(p)(1-\alpha_K \frac{\sqrt{2}\,|n|}{\beta\Jeff})}\frac{1}{|n|(n^2-1)}
f_n(\YG{\tau}_{12}) f_n(\YG{\tau}_{34}) \nn\\
&\simeq \frac{16\Jeff}{\sqrt{2}\alpha_K} \sum_{n}\frac{e^{-i\omega_n\tau}}{|n|(n^2-1)}  \frac{1}{\frac{2\pi|n|}{\beta}+Dp^2} f_n(\YG{\tau}_{12}) f_n(\YG{\tau}_{34}).
\label{eqn: h=2 finite momentum}
\end{align}
Here $\omega_n = \frac{2\pi}{\beta} n$ is the Bosonic Matsubara frequencies. 
In the second step we have expanded $s(p)$ in the long wavelength limit 
${s}(p)\simeq 1-\frac{\calJ_1^2}{3\Jeff^2}p^2$, and defined a constant $D$ which, as will be shown below, is the energy diffusion constant of the system:
\begin{eqnarray}
D=\frac{2\pi \calJ_1^2}{3\sqrt{2}\Jeff \alpha_K}.\label{eqn: diffusionconstant}
\end{eqnarray}
Notice that Eq.~(\ref{eqn: h=2 finite momentum}) has a smooth $p\rightarrow 0$ limit, which reduces to the corresponding formula of the SYK model in Ref.~\cite{maldacena2016comments}.

\begin{figure}
[h]
\center
\subfloat[OPE from the conformal part]
{
\begin{tikzpicture}[scale=0.7,baseline={(current bounding box.center)}]
\filldraw[fill=black] (-40pt,0pt) circle (1pt);
\filldraw[fill=black] (40pt,0pt) circle (1pt);
\draw(-40pt,0pt)--(-70pt,40pt) node[left]{$\chi_j$}
;
\draw(-40pt,0pt)--(-70pt,-40pt)node[left]{$\chi_j$}
 ;
\draw(40pt,0pt)--(70pt,40pt) node[right]{$\chi_k$}
;
\draw(40pt,0pt)--(70pt,-40pt) node[right]{$\chi_k$}
;
\draw [dashed] (-40pt,0pt)--(40pt,0pt);
\node at (0pt,14pt){\scriptsize $\sim c_m^2(p)$};
\end{tikzpicture}
}
\hspace{50pt}
\subfloat[OPE from the correction of $h=2$ parts]
{
\begin{tikzpicture}[scale=0.7,baseline={(current bounding box.center)}]
\filldraw[fill=black] (-40pt,0pt) circle (1pt);
\filldraw[fill=black] (40pt,0pt) circle (1pt);
\draw(-40pt,0pt)--(-70pt,40pt) node[left]{$\chi_j$}
;
\draw(-40pt,0pt)--(-70pt,-40pt)node[left]{$\chi_j$}
 ;
\draw(40pt,0pt)--(70pt,40pt) node[right]{$\chi_k$}
;
\draw(40pt,0pt)--(70pt,-40pt) node[right]{$\chi_k$}
;
\draw [decorate, decoration=zigzag] (-40pt,0pt)--(40pt,0pt);
\node at (0pt,14pt){\scriptsize $\sim \frac{1}{-i \omega + Dp^2}$};
\end{tikzpicture}
}
\caption{ (a) The conformal limit of the four-point function corresponds to collective fields that are locally critical and short-range correlated in space (see equation~(\ref{OPEofconformal})).  (b) The non-conformal corrections of the four-point function corresponds to the reparameterization field, which has a diffusive dynamics in space-time.}
\label{fig: OPE diagram}
\end{figure}
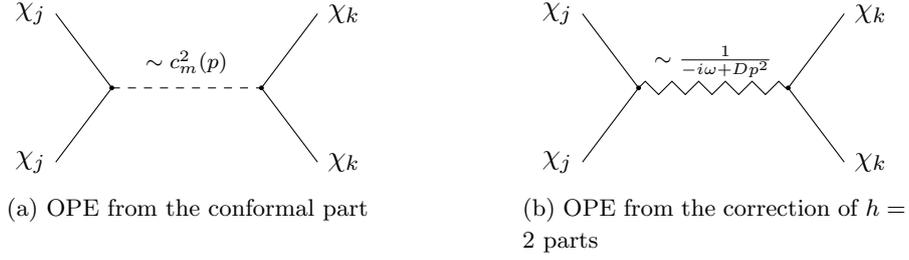

\subsection{Energy transport}

In subsections \ref{subsection: conformal} and \ref{subsection: h=2} we discussed contributions to the fermion four-point functions that are naturally organized in terms of the OPE. 
The operator product of two fermion fields $\chi_{j,x}(\tau+\frac{\YG{\tau}_{12}}2),\chi_{j,x}(\tau-\frac{\YG{\tau}_{12}}2)$ generates an infinite family of collective fields, including the reparameterization field ($h=2$ contribution) and the local conformal fields $\phi_m$. In the strong coupling limit $\beta\Jeff\gg1$, the dominant contribution for small $p$ is the $h=2$ piece, which has diffusive behavior as can be seen from Eq.~(\ref{eqn: h=2 finite momentum}) and (\ref{eqn: reparametrizationaction}), also see figure~\ref{fig: OPE diagram} for an illustration. It is natural to relate the $h=2$ contribution to energy transport properties, since the time reparameterization field is related to diffeomorphisms. 

To understand the relation to energy transport more explicitly, we start from $p=0$, where the $h=2$ four-point function (\ref{eqn: h=2 finite momentum}) reduces to the SYK result. As shown in Ref. \cite{maldacena2016comments}, $\calF_{p=0,h=2}(\tau,\YG{\tau}_{12},\YG{\tau}_{34})$ has a factorized form in the j-j-k-k order region ($\tau_1\YG{>}\tau_2\YG{>}\tau_3\YG{>}\tau_4$), which satisfies the following equality\cite{maldacena2016conformal}:
\begin{equation}
\DS{\frac{1}{N}}\calF_{p=0,h=2}\left(\tau,\tau_{12},\tau_{34}\right)= \frac{\beta^2}{N c_v}  \left( \frac{\partial G^s(\tau_{12})}{\partial \beta } \right)
\left( \frac{\partial G^s(\tau_{34})}{\partial \beta } \right).\label{eqn: energyfluctuation}
\end{equation}
Here $c_v$ is the specific heat per fermion, such that 
\begin{eqnarray}
\frac{NMc_v}{\beta^2}=-\frac{\partial\langle H\rangle}{\partial \beta}=\langle H^2\rangle -\langle H\rangle ^2\equiv \langle H^2\rangle_{\YG{\rm conn.}}
\end{eqnarray} 
is the thermal fluctuation of the total energy. Here $H$ is the Hamiltonian of the chain model. Thus Eq.~(\ref{eqn: energyfluctuation}) can be rewritten as
\begin{eqnarray}
\DS{\frac{1}{N}}\calF_{p=0,h=2}\left(\tau,\tau_{12},\tau_{34}\right)&=&M
\frac{\left( {\partial G^s(\tau_{12})}/{\partial \beta } \right)
\left( {\partial G^s(\tau_{34})}/{\partial \beta } \right)}{\left(\partial\langle H\rangle/\partial\beta\right)^2}\langle H^2\rangle_{\YG{\rm conn.}} \nonumber\\
&=&M\frac{\partial G^s(\tau_{12})}{\partial \langle H\rangle}\frac{\partial G^s(\tau_{34})}{\partial \langle H\rangle} \langle H^2\rangle_{\YG{\rm conn.}} \label{eqn: energyfluctuation2}
\end{eqnarray}

Eq.~(\ref{eqn: energyfluctuation2}) implies that the $h=2$ piece of the fermion four-point function at $p=0$ describes the thermal fluctuation of the two-point function $G^s(\tau_{12})$ induced by the fluctuation of total energy. If we define the energy density of the chain model as $T^{00}(x)$, and define its Fourier component as $T^{00}(p)=M^{-1/2}\sum_xT^{00}(x)e^{-ipx}$, we have $T^{00}(p=0)=H/\sqrt{M}$. It is natural to generalize Eq.~(\ref{eqn: energyfluctuation2}) to finite (small) momentum and express the energy density correlation function as
\begin{align}
\langle \calT_\tau T^{00}(-p,\tau)T^{00}(p,0)\rangle_{\YG{\rm conn.}}\simeq \frac{\calF_{p,h=2}(\tau,\tau_{12},\tau_{34})}{\YG{N}  M^2\frac{\partial G^s(\tau_{12})}{\partial \langle H\rangle}\frac{\partial G^s(\tau_{34})}{\partial \langle H\rangle}}=\frac{Nc_v^2}{\beta^4} \frac{\calF_{p,h=2}(\tau,\tau_{12},\tau_{34})}{ \frac{\partial G^s(\tau_{12})}{\partial\beta } \frac{\partial G^s(\tau_{34})}{\partial \beta }}.\label{eqn: TTand4pt}
\end{align}
Eq.~(\ref{eqn: TTand4pt}) is expected to hold in the limit $\tau_{12},\tau_{34}\rightarrow 0$ (which guarantees the j-j-k-k order for any finite $\tau$). In this limit, by a Taylor expansion of $\calF_{p,h=2}(\tau,\tau_{12},\tau_{34})$ in $\tau_{12}$ and $\tau_{34}$ one obtains
\begin{align}
\frac{\calF_{p,h=2}(\tau,\tau_{12},\tau_{34})}{G^s(\tau_{12})G^s(\tau_{34})}
\simeq \frac{\Jeff}{9\sqrt{2}\alpha_K}\sum_{n}  \frac{ |n| (n^2-1)  e^{-i\omega_n\tau}}{\frac{2\pi|n|}{\beta}+Dp^2} \left(\frac{2\pi\tau_{12}}{\beta}\right)^2\left(\frac{2\pi\tau_{34}}{\beta}\right)^2\nonumber.
\end{align}
Evaluating $\partial_\beta G(\tau)$ for small $\tau$ and using $c_v  =  \frac{\pi \alpha_K}{16\sqrt{2}\beta\calJ}$ and Eq.~(\ref{eqn: TTand4pt}), we find
\begin{eqnarray}
\langle \calT_\tau T^{00}(-p,\tau)T^{00}(p,0)\rangle_{\YG{\rm conn.}}=\frac{Nc_v}{\beta^2} \sum_n  \frac{ |\omega_n |\left(\frac{\beta^2\omega_n^2}{4\pi^2}- 1\right)}{|\omega_n|+Dp^2} e^{-i \omega_n \tau}
\label{eqn: fourier coefficient}
\end{eqnarray}
with $\omega_n=\frac{2\pi n}{\beta}$. This equation directly gives the Matsubara correlator $C^\tau_{00}(p,i\omega_n)$. By analytically continuing
\footnote{One may concern that the fourier coefficient in Eq.~\ref{eqn: fourier coefficient} looks divergent at large $|\omega|$ which makes the analytic continuation less rigorous. However, the expression in the denominator only keeps the leading terms in $(|\omega_n|/J)$ expansion. In general, it includes all higher power terms, i.e. $Dp^2+ |\omega_n| + \sum_{i\geq 2} c_i (|\omega_n|/J)^{-h_i}$ where $c_i$ are coefficients that can be determined by numerical calculations, and $h_i$ are the negative roots of eigenvalue equation $k(h)=1$ (see Eq.~\ref{confeigs} and Ref.~\cite{maldacena2016comments, kitaev2015simple}). We start from $i=2$ in the summation since the first negative root $h_1=-1$ corresponds to the $|\omega_n|$ term. For $i \geq 2$, the roots are generally non-integers with spacing roughly 2: $h=  -2.77354, -4.67946, -6.63197, \ldots$. These higher power terms regulate the large $|\omega|$ behavior for the coefficients, but won't contribute the low energy physics discussed later.}
 $i\omega_n\rightarrow \omega+i\delta$ in $C^\tau_{00}(p,i\omega_n)$ and subtracting a contact term\cite{policastro2003ads}, one obtains the retarded energy-density correlation function in the $(p,\omega)$ space\footnote{We would like to thank Subir Sachdev for pointing out an error in our earlier version.}:
\begin{eqnarray}
C^R_{00}(p,\omega)=  \frac{N c_v}{\beta}    \left( \frac{ -i\omega}{-i\omega+Dp^2}	- 1 \right) +O(\omega^3)\simeq -  \frac{N c_v}{\beta}     \frac{ Dp^2}{-i\omega+Dp^2}.	 \label{eqn: energy diffusion}
\end{eqnarray}
Here we have taken the small frequency limit and omitted the higher order terms in $\omega$. 

Eq.~(\ref{eqn: energy diffusion}) tells us that energy density perturbations in our system satisfy a diffusion equation with diffusion constant $D$ given by Eq.~(\ref{eqn: diffusionconstant}). From energy density correlations one can also derive the thermal conductance:
\begin{equation}
\kappa'(\omega,p) = \Re \left( \frac{i \omega \beta }{p^2 } C^R_{00} (\omega,p)\right) =   \frac{ N c_v D\omega^2}{\omega^2+(Dp^2)^2	}.
\end{equation}
For $p=0$ this formula reduces to $\kappa'=N c_v D=C_vD$, with $C_v=Nc_v$ the specific heat per site. 

\vspace{0.2in}

In summary, in this section we analyzed the fermion four-point function in the limit $\tau_{12},\tau_{34}\rightarrow 0$, which determines the behavior of $\SO(N)$ singlet collective fields in the chain model. The only collective field with nontrivial spatial dynamics is the time reparameterization field, the dynamics of which describes diffusion of energy in this disordered system, with a temperature independent diffusion constant. Although there are infinite number of other collective fields in this system, all of them are locally critical and decay exponentially in space with order $1$ correlation length.

\section{Chaos and the butterfly velocity}
\label{section: chaos and the butterfly velocity}

Another interesting aspect of the model is the OTOC, which have been studied as a diagnostic of chaos. In addition to the exponent $\lambda_L$ that determines the exponential growth of (anti)commutators in the large $N$ dynamics within a single site, the spatial locality in our generalized model allows us to study the dynamics of chaos in space. An important parameter is the speed $v_B$ \cite{shenker2013black,roberts2014localized} defined as the speed of growth of the ``filled light cone'' that marks the region where operators have large (anti)commutators with an initial operator. It has the interpretation of the speed at which the butterfly effect spreads in space, and has also been related to the Lieb-Robinson velocity \cite{Lieb:1972wy} recently\cite{roberts2016lieb,huang2016out}. In this section we will evaluate $v_B$.

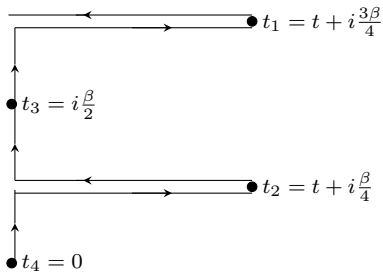
\begin{figure}
[h]
\center
\begin{tikzpicture}[scale=0.6,baseline={(current bounding box.center)}]
\filldraw (0pt,-100pt) circle (3pt) node[right]{\scriptsize $t_4=0$};
\draw[->,>=stealth] (2pt,-100pt)--(2pt,-75pt);
\draw (2pt,-54pt)--(2pt,-75pt);
\draw[->,>=stealth] (2pt,-56pt)--(100pt,-56pt);
\draw (150pt,-56pt)--(75pt,-56pt);
\draw[-<,>=stealth] (2pt,-48pt)--(50pt,-48pt);
\draw (150pt,-48pt)--(40pt,-48pt);

\filldraw (150pt,-52pt) circle (3pt) node[right]{\scriptsize $t_2=t+i\frac{\beta}{4}$};

\draw[->,>=stealth] (2pt,-48pt)--(2pt,-25pt);
\draw (2pt,0pt)--(2pt,-25pt);

\filldraw (0pt,0pt) circle (3pt) node[right]{\scriptsize $ t_3=i\frac{\beta}{2}$};
\draw[->,>=stealth] (2pt,0pt)--(2pt,25pt);
\draw (2pt,48pt)--(2pt,25pt);

\draw[->,>=stealth] (2pt,48pt)--(100pt,48pt);
\draw (150pt,48pt)--(75pt,48pt);
\draw[-<,>=stealth] (-2pt,56pt)--(50pt,56pt);
\draw (150pt,56pt)--(40pt,56pt);

\filldraw (150pt,52pt) circle (3pt) node[right]{\scriptsize $t_1=t+i\frac{3\beta}{4}$};
\end{tikzpicture}
\caption{\YG {Double Keldysh-Schwinger contour with operators equally spaced in imaginary time. }\XLQ{(Note that here we use the convention that the real/imaginary part of $t$ is the real/imaginary time, which is different from our convention in previous section.)}
}

\label{fig: double Keldysh-Schwinger contour}
\end{figure}

The OTOC can be computed by analytic continuation of the imaginary time ordered correlation function. A convenient special case to study is the ``regularized OTOC,'' where the operators are equally spaced in imaginary time \YG{as shown in figure}~\ref{fig: double Keldysh-Schwinger contour}: 
\begin{eqnarray}
F(x,t)&=&\frac{1}{N^2}\sum_{j,k=1}^N \left\langle\chi_{j,x} (t+i\frac{3\beta}4) \chi_{k,0}(i\frac{\beta}2)  \chi_{j,x}(t+i\frac{\beta}4) \chi_{k,0}(0) \right\rangle_\beta
\nonumber\\
&=&\frac{1}{N^2}\sum_{j,k=1}^N  \Tr \left( r \chi_{j,x} (t) r \chi_{k,0}(0) r \chi_{j,x}(t) r \chi_{k,0}(0) \right),\label{OTOC}
\end{eqnarray} 
with $r=\rho(\beta)^{1/4}= e^{-\frac{\beta}4 H}/Z^{1/4}$. The regularization does not change the qualitative behavior of the OTOC. For fixed $x$ we expect the following behavior for $F(x,t)$: at early times it should be  approximately equal to the disconnected product $-G(\frac{\beta}{2})^2$ (the minus sign is because we have fermions), and at late times it should be small, indicating a large anticommutator between $\chi_{j,x}(t)$ and $\chi_{\YG{k,0}}(0)$. The butterfly velocity $v_B$ is defined as the rate at which the region where $F$ is small expands outwards as we increase $t$. Of course, to actually see that $F$ becomes small, we would have to sum all orders in the $\frac{1}{N}$ expansion. We can only compute the first $\frac{1}{N}$ term, but we assume that the exact $F$ becomes small at around the time where this correction becomes comparable to the order one disconnected contribution $-G(\frac{\beta}{2})^2$.

In the imaginary time configuration, the $\frac{1}{N}$ part of the four point function is given by $-\frac{\calF}{N}$, where $\calF$ is the function studied in the previous section. More precisely, we studied the spatial fourier transform $\calF_p$. To compute the $\frac{1}{N}$ term in $F(x,t)$, we will continue $\calF_p$ to the configuration in Eq.~(\ref{OTOC}), and then finally fourier transform back to position space. We will see that there are some subtleties involved in getting an expression for $\calF_p$ that is accurate for the small momenta that dominate this fourier transform. 

To begin, we will warm up by studying the case where $p^2 \gg \frac{\Jeff}{\beta \Jeff_1^2}$ so that we can use the conformal limit of the kernel. From Eq.~(\ref{crossRatioDef}), we have that the cross ratio of the times in the configuration (\ref{OTOC}) is
\begin{equation}\label{cross}
\eta = \frac{2}{1 - i \sinh\frac{2\pi t}{\beta}}.
\end{equation}
For large $t$, this is small and purely imaginary, but we begin the continuation from $t = 0$, where $\eta$ is greater than one. The four point function and this continuation are discussed in detail for the SYK model in Ref.~\cite{maldacena2016comments}. The only difference in our case is that we have to insert a factor of $s(p)$. After the contour manipulation and expansion of the relevant hypergeometric functions \cite{maldacena2016comments}, one finds that the growing term at small $\eta$ is
\begin{align}\label{fpu}
\frac{\mathcal{F}_p(\eta)}{G(\tau_{12})G(\tau_{34})} \sim -\frac{4}{3}\frac{(h{-}1/2)}{\tan(\pi h/2)}\frac{k_R(1-h)}{s({}p)k_R'(1-h)}\frac{\Gamma(\frac{1}{2} - \frac{h}{2})\Gamma(h - \frac{1}{2})}{2^{1-h}\Gamma(\frac{h}{2})}(-i\eta)^{1-h}\Bigg|_{h = h_*(p)}\\
k_R(1-h) = k_0(h)\frac{\cos\pi(\frac{1}{4} - \frac{h}{2})}{\cos\pi(\frac{1}{4}+ \frac{h}{2})} = \frac{3}{2h-1}, \hspace{20pt} h_*(p) = \frac{1 + 3s(p)}{2}.
\end{align}
Using Eq.~(\ref{cross}), we see that this implies an exponential growth $\calF_p\sim e^{\frac{2\pi}{\beta}(h_*(p)-1)t}$, so we have a momentum-dependent chaos exponent $\lambda_L(p) = \frac{2\pi}{\beta}[h_*(p)-1]$.

This exponent is largest at small $p$, where we have $h_*(p) \approx 2 - \frac{\Jeff_1^2}{2\Jeff^2}p^2$. This leads to a $p^{-2}$ divergence in the prefactor coming from the  $\tan(\pi h/2)$ in the denominator of (\ref{fpu}). As before, this divergence can be traced to the $h = 2$ eigenfunctions, which we can treat more accurately by directly continuing Eq.~(\ref{eqn: h=2 finite momentum}) to the configuration (\ref{OTOC}). In appendix \ref{appendix: summation trick}, we show that the growing term after this continuation is
\begin{equation}\label{h=2chaos}
\frac{\calF_{p,h=2}(t)}{G(\frac{\beta}{2})^2} \sim -\frac{\Jeff}{\alpha}\cdot\frac{e^{\frac{2\pi}{\beta}t}}{\frac{2\pi}{\beta} + Dp^2}, \hspace{20pt} \alpha\equiv \frac{\sqrt{2}\alpha_K}{4\pi}.
\end{equation}
This expression has a smooth $p\rightarrow 0$ limit. However, notice that the exponential growth is $e^{\frac{2\pi}{\beta}t}$ independent of $p$, whereas we have argued that at finite $p$ the exact answer should be modified to $e^{\frac{2\pi}{\beta}(h_*(p)-1)t}$. Also, at finite $\beta \Jeff$, we expect a modification of the growth exponent proportional to $\frac{1}{\beta\Jeff}$. 

Both of these modifications must come from the sum over $h\neq 2$ eigenfunctions. We can perturbatively compute both of them as follows. We have two small parameters, $p^2$ and $\frac{1}{\beta\Jeff}$, and we consider both to be first order quantities, of order $\epsilon$. If we expand in $\epsilon$, we expect to find
\begin{equation}
\frac{\calF_p(t)}{G(\frac{\beta}{2})^2} \sim -\frac{1}{B(\epsilon)}e^{\lambda_L(\epsilon)t}, \hspace{20pt} B = b_1\epsilon + b_2\epsilon^2 + ..., \hspace{20pt} \lambda_L = \frac{2\pi}{\beta}(1 - \lambda_1\epsilon + ...).
\end{equation}
At order $\epsilon^{-1}$, we have $\frac{1}{\epsilon b_1}e^{\frac{2\pi}{\beta}t}$. Comparison with (\ref{h=2chaos}) then determines $b_1$. At order $\epsilon^0$ we expect a term $\frac{2\pi\lambda_1}{\beta b_1}t e^{\frac{2\pi}{\beta}t}$ coming from the small shift in $\lambda_L$. Since this is at order $\epsilon^0$, we can compute it in the theory with $p = 0$ and $\beta\Jeff = \infty$. Indeed, Ref.~\cite{maldacena2016comments} did find such a term in this limit, which for our case evaluates to $\frac{6\pi}{\beta}te^{\frac{2\pi}{\beta}t}$. Matching to the expected term, we find $\lambda_1 = 3b_1$, giving the formula
\begin{equation}\label{improved}
\frac{\calF_p(t)}{G(\frac{\beta}{2})^2} \sim - \frac{1}{b(p)}e^{\frac{2\pi}{\beta}[1 - 3b(p)]t}, \hspace{20pt} b(p)=\frac{\alpha}{\Jeff}\left(\frac{2\pi}{\beta} + D p^2\right).
\end{equation}
The exponent in Eq.~(\ref{improved}) is correct to order $p^2$ and order $\frac{1}{\beta\Jeff}$ and the inverse of the prefactor is correct to the same accuracy. One can check that when $\beta\Jeff = \infty$ it agrees with the exact result Eq.~(\ref{fpu}) to the claimed order in $p^2$.

We now have an expression that is sufficiently accurate at small $p$, so we can do the final step and transform Eq.~(\ref{improved}) to position space to compute $F(x,t)$. Approximating the discrete fourier transform as an integral, we have
\begin{equation}
\label{fourier}
\frac{F(x,t)}{-G(\frac{\beta}{2})^2} = 1 - \frac{1}{N}\int_{-\infty}^\infty \frac{dp}{2\pi}\frac{e^{ipx}}{b(p)}e^{\frac{2\pi}{\beta}[1 - 3b(p)]t} + ...
\end{equation}
The integrand has a pole at momentum $p = i\sqrt{\frac{2\pi}{\beta D}}$, which dominates the behavior for large $x$, leading to\footnote{Some constants that appear in the following equations are\[a_1 = \left(\frac{3\beta \Jeff}{\sqrt{2}\alpha_K}\right)^{\frac{1}{2}}\frac{\Jeff}{\Jeff_1}, \hspace{20pt} a_2 = \frac{\beta^\frac{3}{2}\Jeff^2}{\sqrt{2}\pi\alpha_K \Jeff_1}, \hspace{20pt}a_3 = \frac{3\alpha_K}{\sqrt{2}\beta\Jeff}, \hspace{20pt} a_4 = \frac{\beta \Jeff^2}{4\pi\Jeff_1^2}.\]}
\begin{equation}\label{expregion}
\frac{F(x,t)}{-G(\frac{\beta}{2})^2} \simeq 1 - \frac{a_1}{N}e^{\frac{2\pi}{\beta}(t - x/v_B)}, \hspace{20pt} v_B^2 = \frac{2\pi D}{\beta}.
\end{equation}
This is the main result of this section, giving the butterfly velocity $v_B$. The order $\frac{1}{N}$ term competes with the order one term, indicating that the anticommutator has become large, when $t = \frac{\beta}{2\pi}\log\frac{N}{a_1} + \frac{x}{v_B}$. The formula for $v_B$ given in Eq.~(\ref{expregion}) agrees with the relation identified in holographic theories, see Ref.~\cite{blake2016universal}. 

\DS{It is remarkable to find a simple relation between the diffusion constant and butterfly velocity of this type. In this model, it is a consequence of the fact that the same reparameterization degrees of freedom are responsible both for energy diffusion dynamics and the OTOC chaos behavior. This is a property that the model shares with conventional holographic theories, where the gravitational field in the bulk determines both of these phenomena. It would be interesting to work out whether $v_B^2 = \frac{2\pi D}{\beta}$ continues to hold at higher orders in $\frac{1}{\beta J}$. At least naively, because other modes besides reparameterizations will become important, one would not expect this equality to persist beyond infinite $\beta J$.}

\begin{figure}[t]
\center
\includegraphics[scale = 0.4]{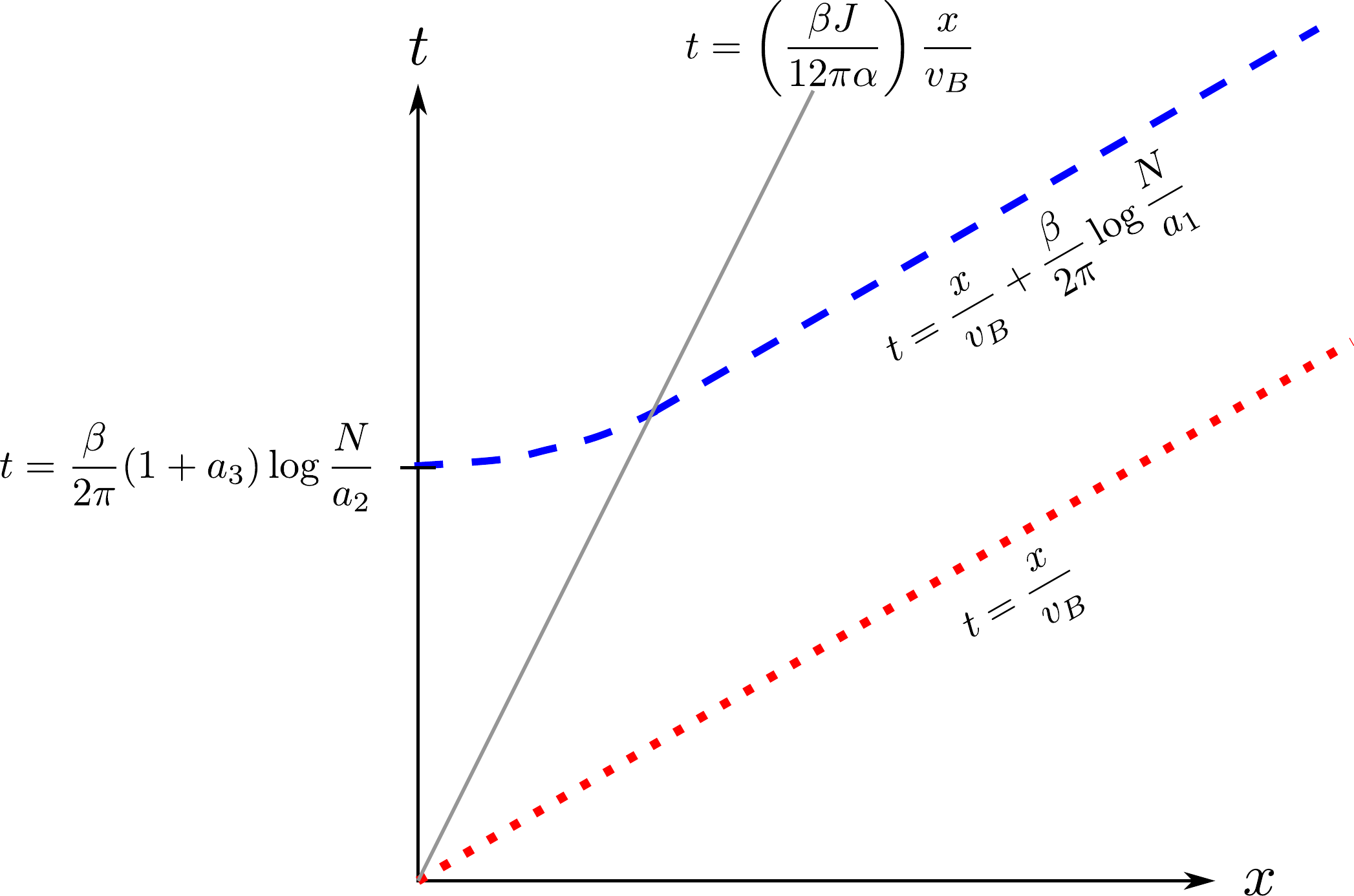}
\caption{We sketch two level sets of the function $F(x,t)$. The red/dotted line corresponds to $F = 1 - \frac{1}{N}$, and the blue/dashed line corresponds to $F = \frac{1}{2}$. The blue/dashed curve is the butterfly cone: operators above this location have large anticommutators with the operator at the origin. The gray solid line marks the transition between the behavior (\ref{expregion}) to the right, and (\ref{diffusionregion}) to the left.}\label{figCone}
\end{figure}

There is an interesting subtlety in the fourier transform that we glossed over above. In fact, the pole dominates only for $x \gg \frac{v_Bt}{\beta \Jeff}$. This means that for $x\lesssim \frac{v_B}{\Jeff}\log N$, the pole analysis will not be correct at the time when the anticommutator becomes large. For such $x$, we can approximate (\ref{fourier}) another way by replacing $b(p)$ in the denominator by $b(0)$ and doing the Gaussian integral. This leads to
\begin{equation}\label{diffusionregion}
\frac{F(x,t)}{-G(\frac{\beta}{2})^2} \simeq 1 - \frac{a_2}{N\sqrt{t}}e^{\frac{2\pi}{\beta}(1-a_3)t - a_4 x^2/t}.
\end{equation}
which is accurate for $x\ll \frac{v_B t}{\beta\Jeff}$. In this region, we find that the ``butterfly cone'' is rounded out, see figure \ref{figCone}. It is rather striking that these two different regions of behavior characterized by (\ref{diffusionregion}) and (\ref{expregion}) were also identified in the analysis of stringy corrections to the holographic computation of $F(x,t)$ in \cite{shenker2014stringy}. 

We will close this section with one further comment. A surprising feature of the large $x$ behavior (\ref{expregion}) is that the growth as a function of time is given by $e^{\frac{2\pi}{\beta}t}$, despite the finite $p$ and $\frac{1}{\beta\Jeff}$ corrections to $\lambda_L$ in Eq.~(\ref{improved}). These corrections, which are both negative for real momenta, cancel against each other when we evaluate at the pole at imaginary $p$. It would be interesting to know if this persists at higher orders in $\frac{1}{\beta\Jeff}$. Note that in the small $x$ region (\ref{diffusionregion}), the growth as a function of time is decreased by a $\frac{1}{\beta\Jeff}$ correction, with the same coefficient as in the original SYK model.

\section{General construction and higher dimensions}
\label{section: general models at higher dimensions}

In previous sections, we have focused on a $(1+1)$-dimensional chain example of the generalized SYK model, but our construction actually applies to generic dimensions and more general forms of interactions.  In this section, we will discuss the general form of our model. We will start from the simple case of higher dimensional regular lattices and then discuss the even more general cases beyond that.

\subsection{Generalized SYK model on higher dimensional lattices}
\begin{figure}
\center
\begin{tikzpicture}[scale=2, baseline={(current bounding box.center)}]
\draw (-20pt,-30pt)--(-20pt,30pt);
\draw (0pt,-30pt)--(-0pt,30pt);
\draw (20pt,-30pt)--(20pt,30pt);

\draw (-30pt,-20pt)--(30pt,-20pt);
\draw (-30pt,0pt)--(30pt,0pt);
\draw (-30pt,20pt)--(30pt,20pt);

\draw[dashed] (-20pt,0pt) --(-35pt,10pt);
\draw[dashed] (-20pt,-20pt) --(-35pt,10pt);
\draw[dashed] (-20pt,20pt) --(-35pt,10pt) node [left]{ $J_{uuwz}$};

\filldraw (-20pt,0pt) circle (1pt) node [below right] {\small $w$};
\filldraw (-20pt,-20pt) circle (1pt) node [below right] {\small $z$};
\filldraw (-20pt,20pt) circle (1pt) node [below right] {\small $u$};
\filldraw (0pt,0pt) circle (1pt);
\filldraw (0pt,-20pt) circle (1pt) node [ below right]{\small $x$};
\filldraw (0pt,20pt) circle (1pt);
\filldraw (20pt,0pt) circle (1pt);
\filldraw (20pt,-20pt) circle (1pt) node [ below right]{\small $y$};
\filldraw (20pt,20pt) circle (1pt);

\draw[dashed] (20pt,-20pt) -- (10pt,-10pt);
\draw[dashed] (0pt,-20pt) -- (10pt,-10pt);
\node at (10pt,-6pt) { $J_{xxyy}$};
\end{tikzpicture}
\caption{An example of two-dimensional square lattice model. The Hamiltonian could contain all kinds of random four-fermion terms, for example, $J_{jklm,uuwz}\chi_{j,u} \chi_{k,u} \chi_{l,w}\chi_{m,z}$ and $J_{jklm,xxyy}\chi_{j,x} \chi_{k,x} \chi_{l,y}\chi_{m,y}$ as shown in the figure.}
\label{fig: generalized model at square lattice}
\end{figure}
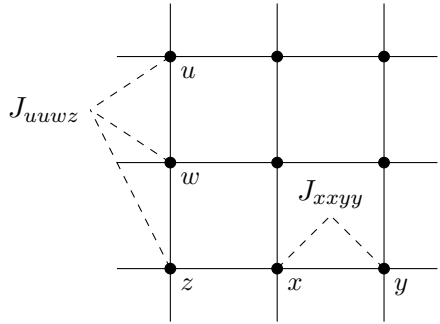

In the chain model example, the different sites are only coupled by a 2-2 coupling with two fermions from each site. This restriction is chosen for simplicity, which is not necessary. In general, our model can be defined on arbitrary graphs, including higher dimensional translation invariant lattices and non-translation invariant graphs \YG{(see figure~(\ref{fig: generalized model at square lattice}) for an illustration on square lattice)}. We denote the set of sites in the graph as $\Gamma$, and label the sites by $x,y,z...$. On each site there are $N$ Majorana fermions $\{\chi_{j,x}\}_{j=1,\ldots,N}$. The fermions are coupled via random four-fermion interactions 
\begin{equation}
H=\sum_{x,y,z,w\in \Gamma}  \sum_{j,l,k,m} J_{jklm,xyzw}\chi_{j,x}\chi_{k,y}\chi_{l,z}\chi_{m,x},\label{eqn:Hgeneral1}
\end{equation}
here \YG{in the sum over $x,y,z,w$, one can always define an order of the sites, and avoid duplication due to different order of $x,y,z,w$. For fixed sites $x,y,z,w$, we can further restrict the range of indices $j,k,l,m$ in the sum to avoid duplication due to the different order of $j,k,l,m$.\footnote{For example: if $x=y=z=w$, we restrict $1\leq j<k<l<m \leq N$; and if $x=y \neq z=w$, we restrict $1\leq j < k \leq N$ and $1\leq l < m \leq N$, as we did in the chain model.} This definition makes sure that each 4-fermion term only appears once in the Hamiltonian.} $\{J_{jklm,xyzw}\}$ are random couplings which are completely independent, with mean and variance
\begin{equation}
\bar{J_{jklm,xyzw}}=0,\quad \bar{J_{jklm,xyzw}^2}=  \frac{1}{N^3} \calJ^2_{xyzw},
\end{equation}
where $\calJ_{xyzw}$ is fixed in the large $N$ limit. This model includes SYK model and the chain model discussed above as two special cases. The SYK model corresponds to taking $\calJ_{xyzw}=\calJ$ uniform for all $x,y,z,w$, which gives a completely non-local Hamiltonian (or equivalently it can be treated as the case $\Gamma$ only contains a single site). The chain model is obtained by setting $\calJ_{xyzw}$ to zero except $\{\calJ_{xxxx}\}_{x\in \Gamma}$ and $\{\calJ_{xxyy}\}_{x\in \Gamma}$ for $y=x \YG{+} 1$.

\begin{figure}
[t]
\center
\subfloat[A watermelon diagram]{
\begin{tikzpicture}[scale=1.5, baseline={(current bounding box.center)}]
\draw (-40pt,0pt)--(40pt,0pt);
\draw (-24pt,0pt)..controls (-10pt,15pt) and (10pt,15pt)..(24pt,0pt);
\draw (-24pt,0pt)..controls (-10pt,-15pt) and (10pt,-15pt)..(24pt,0pt);
\draw[dashed] (-24pt,0pt)..controls (-20pt,30pt) and (20pt,30pt)..(24pt,0pt);
\node at (0pt,3pt) {$z$};
\node at (0pt,15pt) {$y$};
\node at (0pt,-16pt) {$w$};
\node at (0pt,27pt) {$J_{xyzw}$};
\filldraw[fill=black](-40pt,0pt) circle (1pt) node[left]{$x$};
\filldraw[fill=black](40pt,0pt) circle (1pt) node[right]{$x$};
\filldraw[fill=black](-24pt,0pt) circle (1pt);
\filldraw[fill=black](24pt,0pt) circle (1pt);
\end{tikzpicture}
}
\hspace{40pt}
\subfloat[Ladder in generalized model]{
\begin{tikzpicture}[scale=1.5,baseline={([yshift=-4pt]current bounding box.center)}]
\filldraw[fill=black] (-25pt,20pt) circle (1pt) node [left]{$x$};
\filldraw[fill=black] (25pt,20pt) circle (1pt) node [right]{$y$};
\filldraw[fill=black] (-25pt,-20pt) circle (1pt) node [left]{$x$};
\filldraw[fill=black] (25pt,-20pt) circle (1pt) node [right]{$y$};
\draw[thick] (25pt,-20pt)--(-25pt,-20pt);
\draw[thick] (-25pt,20pt)--(25pt,20pt);
\draw[thick] (0pt,20pt)..controls (-10pt,10pt) and (-10pt,-10pt)..(0pt,-20pt);
\draw[thick] (0pt,20pt)..controls (10pt,10pt) and (10pt,-10pt)..(0pt,-20pt);
\draw[dashed] (0pt,20pt)..controls (20pt,10pt) and (20pt,-10pt)..(0pt,-20pt);
\node at (2pt,0pt){$w$};
\node at (-13pt,0pt){$z$};
\node at (25pt,0pt){$J_{xyzw}$};
\end{tikzpicture}}
\caption{(a) A typical ``watermelon'' diagram for the general model. Just like SYK model and the $(1+1)-d$ chain model, in the general model the coupling $J_{xyzw}$ is also diagonal, such that only fermions with the same flavor and spatial coordinate are connected under random average of disorder fields (dashed line). (b) A typical ladder in the generalized model. One needs to sum over all possible $z$ and $w$ in the middle to get the ladder kernel $K_{xy}$.}
\label{fig: general model diagrams}
\end{figure}
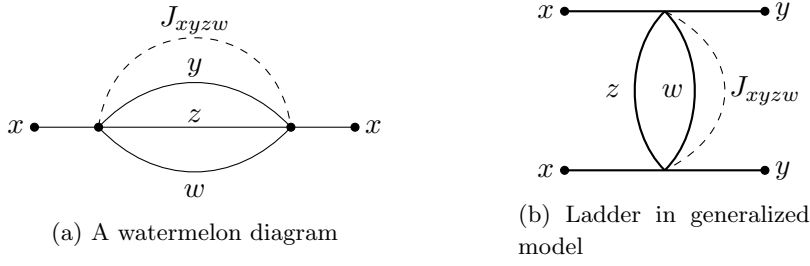

Remarkably, this general model can still be solved in the same way as the chain model. From the Feynman diagrams shown in figure~\ref{fig: general model diagrams}
one can directly see that the disorder averaged two-point function is still diagonal between different sites, even if there is no symmetry reason for it to vanish. In path integral approach, we can write down the effective action of the same collective fields $G_x(\tau_1,\tau_2)$ and $\Sigma_x(\tau_1,\tau_2)$:
\begin{align}
\bar{Z}& = \int \prod_{x\in \Gamma} \calD G_x \calD \Sigma_x \exp(-N S_\eff[G,\Sigma]) \nn \\
S_{\eff}[G,\Sigma] &= \sum_{x\in \Gamma} \left( -\log \Pf \left(\partial_\tau -\Sigma_x  \right)  + \frac{1}{2} \int d\tau_1 d\tau_2 \Sigma_x(\tau_1,\tau_2) G_x(\tau_1,\tau_2) \right), \nn\\
& - \sum_{x,y,z,w\in \Gamma} \left( C_{xyzw} \calJ^2_{xyzw}   \int d\tau_1 d\tau_2    G_x(\tau_1,\tau_2)G_y(\tau_1,\tau_2)G_z(\tau_1,\tau_2)G_w(\tau_1,\tau_2)  \right),
\end{align}
where $C_{xyzw}$ are combinatorial factors which depend on how many sites in $xyzw$ coincide. For example $C_{xxxx}=\YG{\frac{1}{2\cdot 4!}}$, $C_{xxyy}=\YG{\frac{1}{2\cdot 2! \cdot 2!}}$ for  $x\neq y$.\footnote{\XLQ{In general, a group of sites $xyzw$ defines a partition $(n_1,n_2,...,n_k)$ of $4$: $\sum_{i=1}^kn_i=4$ with $n_i$ the multiplicity of site $i$. For example, $xxxx$ corresponds to partition $(4)$ and $xxyy$ corresponds to the partition $(2,2)$. The combinatorial factor is given by $C_{xyzw}= \frac{1}{2} \frac{1}{\prod_i n_i!}$ }} In the large $N$ limit, the corresponding saddle point equation always admits a translation invariant solution $G_x(\tau_1,\tau_2)=G^s(\tau_1,\tau_2)$, which is identical to the two-point function of SYK model with effective coupling 
\begin{eqnarray}
\calJ^2=\frac{8}{|\Gamma|}\sum_{xyzw\in \Gamma} C_{xyzw} \calJ^2_{xyzw}.
\end{eqnarray}
Here we denote $|\Gamma|$ as the total number of sites.

Similar to the chain model, we can expand $G_x(\tau_1,\tau_2)$ around the saddle point by defining $G_x(\tau_1,\tau_2)=G^s(\tau_{12})+|G^s(\tau_{12})|^{-1}g_x(\tau_1,\tau_2)$. Expanding the effective action to quadratic order of $g_x(\tau_1,\tau_2)$ leads to
\begin{align}
\delta S_{\eff}[g] =\YG{ \frac{3J^2}{4}} \sum_{x,y}  \int d^4\tau g_x(\tau_1,\tau_2) \left( \tilde{K}^{-1}(\tau_1,\tau_2;\tau_3,\tau_4)\delta_{xy}-  S_{xy} \delta(\tau_{13})\delta(\tau_{24})\right)  g_y(\tau_3,\tau_4).
\end{align}
The effective action has the same form as equation~\ref{eqn: effective action for fluctuations} except for a different spatial kernel $S_{xy}$:
\begin{eqnarray}
S_{xy}=\YG{\frac{4}{3J^2}} \sum_{z,w\in\Gamma} \YG{(} C_{xyzw} \YG{+C_{xzyw}+C_{xzwy}+C_{zxyw}+C_{zxwy}+C_{zwxy})} \calJ^2_{xyzw}.
\end{eqnarray}

The discussion so far does not rely on translation symmetry, and applies to general graphs. If the graph is a $d$-dimensional lattice and the coupling has translation symmetry, the spatial kernel will also have the same symmetry, so that $S_{xy}=S(x-y)$ where the labels $x,y$ should be considered as $d$-dimensional vectors. In this case the spatial kernel can be diagonalized by Fourier transformation. If $S_{xy}$ is short-ranged, the Fourier transformation $s(\vec{p})$ is a smooth function of $\vec{p}$, which can be expanded at small $p$ as $s(\vec{p})\simeq 1-\sum_{i=1,2,...,d}a_i \YG{p}_i^2$. In the same way as in the chain model, we obtain an energy diffusion, and $a_i$ determines the diffusion constant of energy along the $i$-th direction. 

Following the same approach as in the chain model case, we can also study the OTOC in the general model. Similar to the $(1+1)$-d case, one finds a Lyapunov exponent $\frac{2\pi}\beta$ saturating the chaos bound, and a butterfly velocity $v_B$ (for translation invariant systems). When the spatial kernel $S_{xy}$ is short-ranged, the universal relation $D=\frac{v_B^2}{2\pi T}$ still holds. \XLQ{More details of the higher-dimensional calculation is given in Appendix \ref{appendix: higher dim}.}

\subsection{Models with global symmetry}

As we learned from Ref.~\cite{sachdev1993gapless,sachdev2015bekenstein,fu2016numerical}, a complex fermion version of the SYK model can be defined, which has similar properties such as \YG{local critical} two-point functions. The Hamiltonian of the complex fermion \YG{at chemical potential $\mu=0$} is $H=\sum_{j,k,l,m}J_{jklm}c_j^\dagger c_k^\dagger c_l c_m$ with \YG{$1\leq j <k \leq N$, $1\leq l < m \leq N$}. $J_{jklm}$ are also independent random variables. Since one can always write complex fermion operators in Majorana operators, the complex SYK model can be viewed as a Majorana model with $2N$ Majorana fermions $c_j=\frac12\left(\chi_{j1}+i\chi_{j2}\right)$. The coefficients in this Majorana model are different from that of an SYK model because of the $\UU(1)$ symmetry requirement. 

It is natural to generalize the single-site complex fermion SYK model to models defined on a general graph with a generic global symmetry. The most general form of the Hamiltonian is given by
\begin{equation}
H=\sum_{x,y,z,w\in \Gamma}  \sum_{j,l,k,m=1}^N \sum_{a,b,\YG{c,d;}P}J_{jklm,xyzw}^{P}\eta_P^{abcd}\chi_{j,a,x}\chi_{k,b,y}\chi_{l,c,z}\chi_{m,d,w}\label{eqn:Hgeneral2}
\end{equation}
with $j=1,2,...,N$, $a=1,2,...,L$ labels $NL$ Majorana fermions \YG{$\{ \chi_{j,a,x} \}$} at each site \YG{$x\in \Gamma $}. The index $a$ carries a representation of a global symmetry group. (For Majorana operators the representation has to be real. In other words, $\chi_{i,a,x}$ is transformed under a subgroup of $\SO(L)$.) $\eta_P^{abcd}$ for each $P$ is an invariant rank-$4$ tensor in the symmetry group. For example, for symmetry group $\UU(1)\simeq \SO(2)$, there are two possible invariant tensors $\eta_0^{abcd}=\delta^{ab}\delta^{cd},~\eta_1^{abcd}=\epsilon^{ab}\epsilon^{cd}$. \footnote{A natural way to define $\eta_P^{abcd}$ is to write $\eta_P^{abcd}=f_{ab}^{Pm}f_{cd}^{\bar{P}m}$, with $f_{ab}^{Pm}$ the Clebsch-Gordan coefficients that maps the representation carried by $ab$ to the representation $P$. However, it should be noted that the choice may be redundant. In other words, due to the anti-commutation of Majorana fermion operators, one may not need all representations $P$ to expand an invariant tensor.} We assume the couplings $J_{ijkl}^{P}$ are independent variables with
\begin{eqnarray}
\overline{J_{ijkl,xyzw}^P}=0,\quad \overline{\left(J_{ijkl,xyzw}^{P}\right)^2}=\frac1{N^3} \left({\calJ_{xyzw}^{P}} \right)^2.
\end{eqnarray}

Once the Hamiltonian is defined one can try to study the effective action in the same way as before. Now we have to introduce the matrix bilocal field
\begin{eqnarray}
G_x^{ab}(\tau_1,\tau_2)=\frac1N\sum_j\langle \chi_{j,a,x}(\tau_1)\chi_{j,b,x}(\tau_2)\rangle
\end{eqnarray}
and the corresponding Lagrange multiplier $\Sigma_x^{ab}(\tau_1,\tau_2)$. The effective action after integrating out fermions is
\begin{eqnarray}
S_{\eff}[G,\Sigma] &= &\sum_{x\in \Gamma} \left( -\log \Pf \left(\partial_\tau\delta_{ab} -\Sigma_x^{ab}  \right)  + \frac{1}{2} \int d\tau_1 d\tau_2 \Sigma_x^{ab}(\tau_1,\tau_2) G_x^{ab}(\tau_1,\tau_2) \right) \nn\\
& &-\sum_{x,y,z,w\in \Gamma}\sum_P \left[C^P_{xyzw} \left(\calJ^P_{xyzw}\right)^2 \right.\nonumber\\
& &\cdot\left.  \int d\tau_1 d\tau_2  \eta_{abcd}^P\eta_{a'b'c'd'}^P  G_x^{aa'}(\tau_1,\tau_2)G_y^{bb'}(\tau_1,\tau_2)G_z^{cc'}(\tau_1,\tau_2)G_w^{dd'}(\tau_1,\tau_2)  \right]
\end{eqnarray}

Although the effective action is more complicated, the large $N$ saddle point approximation still apply, at least if we keep $L$ finite in the large $N$ limit. \footnote{The suppression of inter-replica coupling by large $N$ still applies to in this limit, so that we expect the disorder averaged effective action with a single copy is still meaningful.} The saddle point condition of this effective action gives the Schwinger-Dyson equations of $G$ and $\Sigma$. In general there may be saddle points where $G^{ab}$ and $\Sigma^{ab}$ have off-diagonal matrix elements, which we don't know how to solve analytically. However, there always exists a 
diagonal saddle point solution $G^{ab}_x(\tau_1,\tau_2)=\delta^{ab}G^s(\tau_{12})$, for which the effective action reduces to the SYK model with a coupling
\begin{eqnarray}
\calJ^2=\frac{8}{|\Gamma|L}\sum_{x,y,z,w}\sum_PC^{P}_{xyzw} \left(\calJ^P_{xyzw} \right)^2\eta^P_{abcd}\eta^P_{abcd}.
\end{eqnarray}
Therefore one can always take an expansion around this saddle point and study its stability using our knowledge about SYK model solution. If this saddle point is stable, these models have the same locally critical two-point functions as the SYK model, but have different $\frac1N$ fluctuations since the fluctuation $g_x^{ab}(\tau_1,\tau_2)$ is a matrix field. This will lead to different four-point functions. We will leave more systematic investigation of these models to future works.

\subsection{General $q$}

Another type of generalization one can consider is to include interactions that involve $q$ fermions at a time, rather than just four. For $q\geq 4$, the detailed analysis of the chain model will be very similar to $q = 4$, with a few minor changes in numerical coefficients. The relation $v_B^2 = 2\pi D/\beta$ will remain correct at large coupling. One can also consider including terms with different values of $q$ in the same model. In general, the terms with higher values of $q$ are more RG-irrelevant, so the terms with the lowest values of $q$ will dominate in the infrared. Nevertheless, the subleading terms can have interesting effects. Suppose we have terms with $q_1,q_2$, with $q_1<q_2$, and we use the same dimensionful coupling $J$ for both interactions. Then a perturbative analysis of the Schwinger Dyson equations indicates that for $\tau \ll \beta$, $\tau J \gg 1$ we will have (omitting coefficients)
\begin{equation}\label{generalqexp}
G(\tau) \propto \frac{\sgn(\tau)}{|J\tau|^\frac{2}{q_1}} + \frac{\sgn(\tau)}{|J\tau|^{\frac{2}{q_1} + 1}} + \frac{\sgn(\tau)}{|J\tau|^{p}}+... \hspace{20pt} p = \frac{2}{q_1} + 2\left(\frac{q_2}{q_1} - 1\right).
\end{equation}
The first term is the naive conformal limit in the pure $q_1$ theory. The second term is the correction to the conformal limit, again in the pure $q_1$ theory; this term leads to the $\alpha_K$ correction to the kernel that eventually gives the action for the reparameterization modes. The third term is the new feature of the theory with both $q_1$ and $q_2$. It reflects the contribution of the irrelevant $q_2$-fermion operator, which affects the correlator at quadratic order. The dimension of this operator near the IR is $\Delta = \frac{q_2}{q_1}$, and when $\Delta < \frac{3}{2}$ we have $p < \frac{2}{q_1} + 1$, so this third term will dominate over the second in (\ref{generalqexp}). In such a case, the analysis of the reparameterization action would have to be redone, following e.g. appendix D of \cite{maldacena2016conformal}. It would be interesting to compute the energy and chaos dynamics in the resulting model.

\section{Conclusion and discussion}
\label{section: conclusion and discussion}

In this paper, we generalize the $(0+1)$-dimensional SYK model of Majorana fermions to higher spatial dimensions. The generalized model retains many interesting properties of the SYK model, such as local criticality, extensive zero temperature entropy and maximal chaos. On top of that, the spatial locality of our generalized model leads to many new physical properties. We find that single Majorana fermions in our model do not propagate between different sites, but collective modes made by pairs of fermions have nontrivial spatial dynamics. In particular, the most important collective mode in the low-energy-long-wavelength limit is the time reparameterization field, the dynamics of which describes the diffusion of energy in this system, with a temperature independent diffusion constant $D$. This result tells us that our model describes a strongly correlated diffusive metal. The dynamics of the same reparameterization field also determines OTOC of fermion operators, from which we can also study the butterfly effect in this model. Our result shows that chaos spreads in space with a ``butterfly velocity" $v_B$. 
Remarkably, at strong coupling, the diffusion constant and the butterfly velocity satisfies a simple relation $D=v_B^2/2\pi T$, in consistency with the proposal in the literature about incoherent metals\cite{hartnoll2015theory,blake2016universalcharge, blake2016universal}.

Our model pointed out a new class of solvable interacting lattice models in condensed matter physics. Usually, solvable models are mapped to weakly interacting theories such as mean field theories, so that they are not ``chaotic", while the interesting phenomena in chaotic systems cannot be studied in solvable models. The generalizations of SYK model is a rare example of solvable but still chaotic systems. Therefore this model provides an interesting platform for studying various properties of strongly correlated systems, such as thermalization, entanglement propagation, dissipative transport, etc. It is also natural to ask whether further generalizations of these models allow us to investigate the possibility of many-body localization and phase transition between localized and delocalized phases. 

From the perspective of holographic duality, the generalized SYK models might be considered as models that are dual to some kind of incoherent black hole (see\cite{blake2016universal} and references therein), but the details of this duality require further work. \DS{At strong coupling, the models do share a key property with conventional holographic systems, which is that a single set of degrees of freedom dominate and describe both energy diffusion and the chaos behavior. In a holographic theory, the relevant degrees of freedom are the bulk gravitational field. Here, it is a reparameterization of time that can vary from place to place. It would be interesting to derive the action (\ref{eqn: reparametrizationaction}) for these degrees of freedom from a subset of the metric degrees of freedom on some black hole background,} in a similar way as the derivation of (0+1)-dimensional Schwarzian action from the Einstein-dilaton theory in approximately AdS$_2$ background \cite{almheiri2014models,maldacena2016conformal}. It would also be interesting to understand the relationship of this action to recent work on hydrodynamic actions \cite{haehl2015adiabatic,crossley2015off}. Another natural question is whether there are higher-dimensional translation invariant generalizations of SYK models which are dual to weakly coupled gravity theories in the bulk.

\section*{Acknowledgment}

We would like to thank 
Mike Blake,
Richard Davison,
Luca V. Delacr\'etaz,
Wenbo Fu,
Tarun Grover,
Sean Hartnoll,
Alexei Kitaev,
Juan Maldacena
and Steve Shenker
for helpful discussions.
We especially acknowledge Subir Sachdev for helpful discussions and comments on the draft. This work is supported by the National Science Foundation through the grant No. DMR-1151786 (YG and XLQ). D.S. is supported by the Simons Foundation grant 385600.

\appendix

\section{Diagrammatic derivation for four-point functions}
\label{subsection: four-point function, diagrams}

In this section, we present a diagrammatic derivation of the fermion four-point functions. The four-point functions are the leading correlation functions which couple different sites. In the language of collective field $G_x(\tau_1,\tau_2)$, the four-point function comes from its quantum fluctuations. The four-point functions are essential for understanding transport properties and also the OTOC measure of chaos, as discussed in section \ref{section: transport} and \ref{section: chaos and the butterfly velocity}.

The connected four-point function is defined as
\begin{align}
\frac{\calF_{xy} (\tau_1,\tau_2; \tau_3,\tau_4)}{N}:= \frac{1}{N^2}\sum_{j,k=1}^N  \langle \calT_\tau \chi_{j,x}(\tau_1) \chi_{j,x}(\tau_2) \chi_{k,y}(\tau_3) \chi_{k,y}(\tau_4) \rangle - G^s_x(\tau_1,\tau_2)G^s_y(\tau_3,\tau_4)
\end{align}
We follow the main text to use $G^s$ to denote the saddle point of bilocal field, which is the Green's function here. Note that the four-point function is non-vanishing after disorder average only if the spatial and flavor indices appear in pairs, \YG{i.e. $\SO(N)$ singlet}. 

\begin{figure}
[t]
\center
\subfloat[$J$-$J$ contraction]{
\begin{tikzpicture}[scale=1.4,baseline={([yshift=-4pt]current bounding box.center)}]
\filldraw[fill=black] (-25pt,20pt) circle (1pt) node [left]{$x$};
\filldraw[fill=black] (25pt,20pt) circle (1pt) node [right]{$x$};
\filldraw[fill=black] (-25pt,-20pt) circle (1pt) node [left]{$x$};
\filldraw[fill=black] (25pt,-20pt) circle (1pt) node [right]{$x$};
\draw[thick] (25pt,-20pt)--(-25pt,-20pt);
\draw[thick] (-25pt,20pt)--(25pt,20pt);
\draw[thick] (0pt,20pt)..controls (-10pt,10pt) and (-10pt,-10pt)..(0pt,-20pt);
\draw[thick] (0pt,20pt)..controls (10pt,10pt) and (10pt,-10pt)..(0pt,-20pt);
\draw[dashed] (0pt,20pt)..controls (20pt,10pt) and (20pt,-10pt)..(0pt,-20pt);
\node at (2pt,0pt){$x$};
\node at (-13pt,0pt){$x$};
\node at (25pt,0pt){$J$};
\end{tikzpicture}}~\quad~
\subfloat[$J'$-$J'$ contraction, $y=x\pm 1$]{
\begin{tikzpicture}[scale=1.4,baseline={([yshift=-4pt]current bounding box.center)}]
\filldraw[fill=black] (-25pt,20pt) circle (1pt) node [left]{$x$};
\filldraw[fill=black] (25pt,20pt) circle (1pt) node [right]{$x$};
\filldraw[fill=black] (-25pt,-20pt) circle (1pt) node [left]{$x$};
\filldraw[fill=black] (25pt,-20pt) circle (1pt) node [right]{$x$};
\draw[thick] (25pt,-20pt)--(-25pt,-20pt);
\draw[thick] (-25pt,20pt)--(25pt,20pt);
\draw[thick] (0pt,20pt)..controls (-10pt,10pt) and (-10pt,-10pt)..(0pt,-20pt);
\draw[thick] (0pt,20pt)..controls (10pt,10pt) and (10pt,-10pt)..(0pt,-20pt);
\draw[dashed] (0pt,20pt)..controls (20pt,10pt) and (20pt,-10pt)..(0pt,-20pt);
\node at (2pt,0pt){$y$};
\node at (-13pt,0pt){$y$};
\node at (25pt,0pt){$J'$};
\end{tikzpicture}}~\quad~
\subfloat[$J'$-$J'$ contraction, $y=x\pm 1$]{
\begin{tikzpicture}[scale=1.4,baseline={([yshift=-4pt]current bounding box.center)}]
\filldraw[fill=black] (-25pt,20pt) circle (1pt) node [left]{$x$};
\filldraw[fill=black] (25pt,20pt) circle (1pt) node [right]{$y$};
\filldraw[fill=black] (-25pt,-20pt) circle (1pt) node [left]{$x$};
\filldraw[fill=black] (25pt,-20pt) circle (1pt) node [right]{$y$};
\draw[thick] (25pt,-20pt)--(-25pt,-20pt);
\draw[thick] (-25pt,20pt)--(25pt,20pt);
\draw[thick] (0pt,20pt)..controls (-10pt,10pt) and (-10pt,-10pt)..(0pt,-20pt);
\draw[thick] (0pt,20pt)..controls (10pt,10pt) and (10pt,-10pt)..(0pt,-20pt);
\draw[dashed] (0pt,20pt)..controls (20pt,10pt) and (20pt,-10pt)..(0pt,-20pt);
\node at (2pt,0pt){$y$};
\node at (-13pt,0pt){$x$};
\node at (25pt,0pt){$J'$};
\end{tikzpicture}}
\caption{Three types of the ``rungs'': type (a) is the same as the one appears in SYK, induced by the interactions between fermions at same sites; type (b) comes from the interactions between fermions at site $x$ and nearest neighbor $y=x\pm1$, but the ``rails'' carry the same site indices, therefore, the effect of interaction doesn't propagate to next site; type (c) also comes from the interactions between nearest neighbor sites, and the rails get shifted by $\pm 1$.}
\label{fig: 3 tpyes of rungs}
\end{figure}
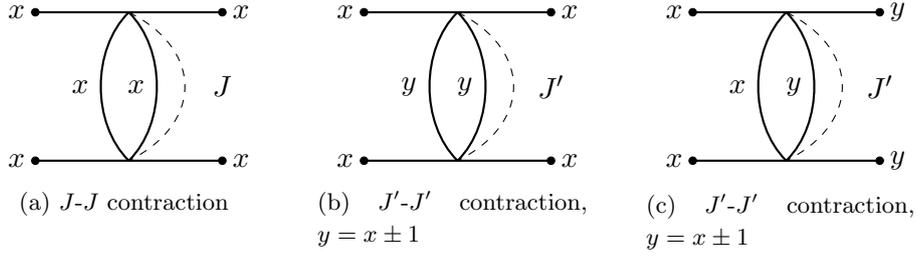

Similar to the SYK model, in the large $N$ limit, the leading contributions to the connected piece of four-point functions are of order $1/N$ and consist of the ladder diagrams:
\begin{align}
\calF\sim \begin{tikzpicture}[scale=1,baseline={([yshift=-4pt]current bounding box.center)}]
\draw[thick] (10pt,10pt) -- (40pt,10pt);
\draw[thick] (10pt,-10pt) --(40pt,-10pt);
\end{tikzpicture}
+
\begin{tikzpicture}[scale=1,baseline={([yshift=-4pt]current bounding box.center)}]
\draw[white] (-30pt,-20pt) rectangle (30pt,20pt); 
\draw[thick] (-30pt,10pt) -- (30pt,10pt);
\draw[thick] (-30pt,-10pt) --(30pt,-10pt);
\draw[thick] (0pt,10pt) .. controls (-5pt,2pt) and (-5pt,-2pt) .. (0pt,-10pt);
\draw[thick] (0pt,10pt) .. controls (5pt,2pt) and (5pt,-2pt) .. (0pt,-10pt);
\end{tikzpicture} + 
\ldots-
\begin{tikzpicture}[scale=1,baseline={([yshift=-4pt]current bounding box.center)}]
\draw[thick] (10pt,10pt) -- (30pt,10pt);
\draw[thick] (10pt,-10pt) --(30pt,-10pt);
\draw[thick] (30pt,10pt) -- (50pt,-10pt);
\draw[line width=5pt,white] (30pt,-10pt) -- (50pt,10pt);
\draw[thick] (30pt,-10pt) -- (50pt,10pt);
\end{tikzpicture}
 -
\begin{tikzpicture}[scale=1,baseline={([yshift=-4pt]current bounding box.center)}]
\draw[thick] (-30pt,10pt) -- (10pt,10pt);
\draw[thick] (10pt,10pt) -- (30pt,-10pt);
\draw[line width=5pt,white] (10pt,-10pt) -- (30pt,10pt);
\draw[thick] (10pt,-10pt) -- (30pt,10pt);
\draw[thick] (-30pt,-10pt) --(10pt,-10pt);
\draw[thick] (0pt,10pt) .. controls (-5pt,2pt) and (-5pt,-2pt) .. (0pt,-10pt);
\draw[thick] (0pt,10pt) .. controls (5pt,2pt) and (5pt,-2pt) .. (0pt,-10pt);
\end{tikzpicture}
 -\ldots
 \label{eqn: sum of ladders}
\end{align}
where the thick lines are dressed Green's functions solved in the previous sections. The interaction vertices are paired by random disorder fields $J_x$ and/or $J'_x$. 
\begin{figure}
[t]
\center
\begin{tikzpicture}[scale=1,baseline={(current bounding box.center)}]
\draw(-60pt,30pt)--(120pt,30pt);
\filldraw[black](-60pt,30pt) circle (1pt) node [left]{$x,\tau_1$};
\filldraw[black](120pt,30pt) circle (1pt) node [right]{$y,\tau_3$};
\draw(-60pt,-30pt)--(120pt,-30pt);
\filldraw[black](-60pt,-30pt) circle (1pt) node [left]{$x,\tau_2$};
\filldraw[black](120pt,-30pt) circle (1pt) node [right]{$y,\tau_4$};
\filldraw[black](-15pt,-30pt) circle (1pt);
\filldraw[black](-15pt,30pt) circle (1pt);
\draw (-15pt,30pt)..controls (-5pt,10pt) and (-5pt,-10pt) ..(-15pt,-30pt);
\draw (-15pt,30pt)..controls (-25pt,10pt) and (-25pt,-10pt) ..(-15pt,-30pt);
\draw[dashed] (-15pt,30pt)..controls (-35pt,10pt) and (-35pt,-10pt) ..(-15pt,-30pt);

\filldraw[black](20pt,-30pt) circle (1pt);
\filldraw[black](20pt,30pt) circle (1pt);
\draw (20pt,30pt)..controls (30pt,10pt) and (30pt,-10pt) ..(20pt,-30pt);
\draw (20pt,30pt)..controls (10pt,10pt) and (10pt,-10pt) ..(20pt,-30pt);
\draw[dashed] (20pt,30pt)..controls (0pt,10pt) and (0pt,-10pt) ..(20pt,-30pt);

\filldraw[black](80pt,-30pt) circle (1pt);
\filldraw[black](80pt,30pt) circle (1pt);
\draw (80pt,30pt)..controls (90pt,10pt) and (90pt,-10pt) ..(80pt,-30pt);
\draw (80pt,30pt)..controls (70pt,10pt) and (70pt,-10pt) ..(80pt,-30pt);
\draw[dashed] (80pt,30pt)..controls (60pt,10pt) and (60pt,-10pt) ..(80pt,-30pt);
\node at (50pt,0pt){$\ldots$};
\end{tikzpicture}
\caption{Ladder diagrams connecting two sites far apart. To connect two sites with distance $n=|x-y|$, one need at least $n$ ladders. And the four-point function includes all possible such diagrams and the partner terms with $(\tau_3\leftrightarrow \tau_4)$.}
\label{fig: boxes with distance n}
\end{figure}
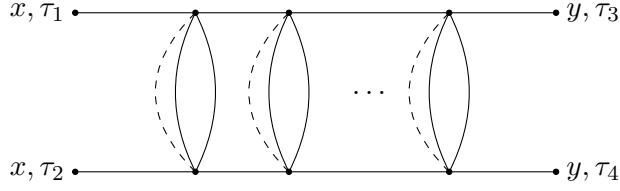
Comparing to those in the SYK model\cite{kitaev2015simple,
polchinski2016spectrum,
maldacena2016comments}, the ladder diagrams in the chain model have extra labels indicating the spatial coordinates (figure~\ref{fig: boxes with distance n}). Each ladder can either couple fermions on the same site, or bring two fermions at site $x$ to the neighboring site $x\pm 1$, as is shown in figure~\ref{fig: 3 tpyes of rungs}. The on-site terms are contributed by both $J_x$ and $J'_x$ terms in the Hamiltonian, while the nearest neighbor terms are only from the $J'_x$ term.

In general, we can complete the summation of ladder diagrams (equation~\ref{eqn: sum of ladders}) using the Schwinger-Dyson equation:
\begin{eqnarray}
\begin{tikzpicture}[scale=1.2,baseline={([yshift=-4pt]current bounding box.center)}]
\draw[thick] (10pt,10pt) -- (50pt,10pt);
\draw[thick] (10pt,-10pt) --(50pt,-10pt);
\filldraw[fill=gray] (20pt,-10pt) rectangle (40pt,10pt);
\end{tikzpicture}=
\begin{tikzpicture}[scale=1.2,baseline={([yshift=-4pt]current bounding box.center)}]
\draw[thick] (10pt,10pt) -- (40pt,10pt);
\draw[thick] (10pt,-10pt) --(40pt,-10pt);
\end{tikzpicture}-
\begin{tikzpicture}[scale=1.2,baseline={([yshift=-4pt]current bounding box.center)}]
\draw[thick] (10pt,10pt) -- (30pt,10pt);
\draw[thick] (10pt,-10pt) --(30pt,-10pt);
\draw[thick] (30pt,10pt) -- (50pt,-10pt);
\draw[line width=5pt,white] (30pt,-10pt) -- (50pt,10pt);
\draw[thick] (30pt,-10pt) -- (50pt,10pt);
\end{tikzpicture} +
\begin{tikzpicture}[scale=1.2,baseline={([yshift=-4pt]current bounding box.center)}]
\draw[thick] (-20pt,10pt) -- (40pt,10pt);
\draw[thick] (-20pt,-10pt) --(40pt,-10pt);
\draw[thick] (0pt,10pt) .. controls (-5pt,2pt) and (-5pt,-2pt) .. (0pt,-10pt);
\draw[thick] (0pt,10pt) .. controls (5pt,2pt) and (5pt,-2pt) .. (0pt,-10pt);
\filldraw[fill=gray] (15pt,-10pt) rectangle (25pt,10pt);
\end{tikzpicture}
\end{eqnarray}
where the gray box represents the dressed interaction vertex. In terms of algebraic formula, this equation is written as
\begin{align}
\calF &= F_0+ K \calF ~\Longrightarrow~
\calF = (1-K)^{-1} F_0\\
\text{with~}
\calF &= \begin{tikzpicture}[scale=1.2,baseline={([yshift=-4pt]current bounding box.center)}]
\draw[thick] (10pt,10pt) -- (50pt,10pt);
\draw[thick] (10pt,-10pt) --(50pt,-10pt);
\filldraw[fill=gray] (20pt,-10pt) rectangle (40pt,10pt);
\end{tikzpicture}, \quad
K = 
\begin{tikzpicture}[scale=1.2,baseline={([yshift=-4pt]current bounding box.center)}]
\draw[thick] (-20pt,10pt) -- (0pt,10pt);
\draw[thick] (-20pt,-10pt) --(0pt,-10pt);
\draw[thick] (0pt,10pt) .. controls (-5pt,2pt) and (-5pt,-2pt) .. (0pt,-10pt);
\draw[thick] (0pt,10pt) .. controls (5pt,2pt) and (5pt,-2pt) .. (0pt,-10pt);
\end{tikzpicture}, \quad
F_0 = \begin{tikzpicture}[scale=1.2,baseline={([yshift=-4pt]current bounding box.center)}]
\draw[thick] (10pt,10pt) -- (40pt,10pt);
\draw[thick] (10pt,-10pt) --(40pt,-10pt);
\end{tikzpicture}-
\begin{tikzpicture}[scale=1.2,baseline={([yshift=-4pt]current bounding box.center)}]
\draw[thick] (10pt,10pt) -- (30pt,10pt);
\draw[thick] (10pt,-10pt) --(30pt,-10pt);
\draw[thick] (30pt,10pt) -- (50pt,-10pt);
\draw[line width=5pt,white] (30pt,-10pt) -- (50pt,10pt);
\draw[thick] (30pt,-10pt) -- (50pt,10pt);
\end{tikzpicture} 
\nonumber
\end{align}
In this equation, the ladder kernel $K=K_{xy}(\tau_1,\tau_2;\tau_3,\tau_4)$ is treated as an operator acting on functions of one spatial coordinate $x$ and two time variables. There are two kinds of nonzero matrix elements of $K$, given by the diagrams in figure~\ref{fig: 3 tpyes of rungs} (a) (b) and those in (c) respectively:
\begin{align}
K_{xx}(\tau_1,\tau_2;\tau_3,\tau_4)&= 
\begin{tikzpicture}[scale=0.8,baseline={([yshift=-4pt]current bounding box.center)}]
\filldraw[fill=black] (-40pt,20pt) circle (1pt) node [left]{$1,x$};
\filldraw[fill=black] (10pt,20pt) circle (1pt) node [right]{$3,x$};
\filldraw[fill=black] (-40pt,-20pt) circle (1pt) node [left]{$2,x$};
\filldraw[fill=black] (10pt,-20pt) circle (1pt) node [right]{$4,x$};
\draw[thick] (10pt,-20pt)--(-40pt,-20pt);
\draw[thick] (-40pt,20pt)--(10pt,20pt);
\draw[thick] (0pt,20pt)..controls (-10pt,10pt) and (-10pt,-10pt)..(0pt,-20pt);
\draw[thick] (0pt,20pt)..controls (10pt,10pt) and (10pt,-10pt)..(0pt,-20pt);
\draw[dashed] (0pt,20pt)..controls (20pt,10pt) and (20pt,-10pt)..(0pt,-20pt);
\node at (25pt,0pt){$J$};
\end{tikzpicture}
~
+
~
\begin{tikzpicture}[scale=0.8,baseline={([yshift=-4pt]current bounding box.center)}]
\filldraw[fill=black] (-40pt,20pt) circle (1pt) node [left]{$1,x$};
\filldraw[fill=black] (10pt,20pt) circle (1pt) node [right]{$3,x$};
\filldraw[fill=black] (-40pt,-20pt) circle (1pt) node [left]{$2,x$};
\filldraw[fill=black] (10pt,-20pt) circle (1pt) node [right]{$4,x$};
\draw[thick] (10pt,-20pt)--(-40pt,-20pt);
\draw[thick] (-40pt,20pt)--(10pt,20pt);
\draw[thick] (0pt,20pt)..controls (-10pt,10pt) and (-10pt,-10pt)..(0pt,-20pt);
\draw[thick] (0pt,20pt)..controls (10pt,10pt) and (10pt,-10pt)..(0pt,-20pt);
\draw[dashed] (0pt,20pt)..controls (20pt,10pt) and (20pt,-10pt)..(0pt,-20pt);
\node at (25pt,0pt){$J'$};
\end{tikzpicture}\nn\\
&=(3\calJ_{0}^2+\calJ_1^2) G^s_x(\tau_{13}) G^s_x(\tau_{34})^2 G^s_x(\tau_{42})=: K_1
\end{align}
\begin{align}
K_{x,x+1}(\tau_1,\tau_2;\tau_3,\tau_4)&= 
\begin{tikzpicture}[scale=0.8,baseline={([yshift=-4pt]current bounding box.center)}]
\filldraw[fill=black] (-40pt,20pt) circle (1pt) node [left]{$1,x$};
\filldraw[fill=black] (10pt,20pt) circle (1pt) node [right]{$3,y=x\pm1$};
\filldraw[fill=black] (-40pt,-20pt) circle (1pt) node [left]{$2,x$};
\filldraw[fill=black] (10pt,-20pt) circle (1pt) node [right]{$4,y=x\pm1$};
\draw[thick] (10pt,-20pt)--(-40pt,-20pt);
\draw[thick] (-40pt,20pt)--(10pt,20pt);
\draw[thick] (0pt,20pt)..controls (-10pt,10pt) and (-10pt,-10pt)..(0pt,-20pt);
\draw[thick] (0pt,20pt)..controls (10pt,10pt) and (10pt,-10pt)..(0pt,-20pt);
\draw[dashed] (0pt,20pt)..controls (20pt,10pt) and (20pt,-10pt)..(0pt,-20pt);
\node at (25pt,0pt){$J'$};
\end{tikzpicture} \nn \\
&=\calJ_1^2 G^s_x(\tau_{13}) G^s_x(\tau_{34}) G^s_y(\tau_{34}) G^s_x(\tau_{42})=: K_2
\end{align}
For the translation invariant saddle point solution, $G^s_x(\tau)$ is independent from $x$, so that $K_1$ and $K_2$ are only different by the coefficient in front. Therefore $K_{xy}(\tau_1,\tau_2;\tau_3,\tau_4)$ has the separable form
\begin{eqnarray}
K_{xy}(\tau_1,\tau_2; \tau_3,\tau_4)&=&S_{xy}K(\tau_1,\tau_2;\tau_3,\tau_4)
\label{eqn: factorized kernel}\\
\text{with~}S_{xy}&=& \delta_{x,y} +\frac{\calJ_1^2 }{3\Jeff^2}\left(\delta_{x,y\pm1} -2\delta_{xy}\right)\\
K(\tau_1,\tau_2;\tau_3,\tau_4)&=& 3\Jeff^2 G^s(\tau_{13}) G^s(\tau_{34})^2 G^s(\tau_{42}) 
\end{eqnarray}
The spatial kernel $S_{xy}$ is a simple tight-binding hopping matrix ({ i.e.} an identity matrix plus a lattice Laplacian), and the temperal kernel $K(\tau_1,\tau_2;\tau_3,\tau_4)$ is identical to that of the $(0+1)$-d SYK model with coupling constant $\Jeff$. (One should be reminded that we denote $\Jeff=\sqrt{\calJ_0^2+\calJ_1^2}$.) The separable form of the kernel $K_{xy}(\tau_1,\tau_2;\tau_3,\tau_4)$ allows us to directly apply the results in the SYK model\cite{kitaev2015simple,maldacena2016comments} to diagonalize the kernel and solve the four-point functions in our model.

Assuming a formal diagonalization $K_{xy}=\sum_{h,n,p} k(h,n,p) | \Psi_{h,n,p} \rangle \langle \Psi_{h,n,p} |$,  
we can express the four-point function by the inner product:
$\calF = \sum_{h,n,p}  | \Psi_{h,n,p} \rangle  \frac{1}{1-k(h,n,p)} \langle \Psi_{h,n,p} | F_0 \rangle$
where $|\Psi_{h,n,p} \rangle = \Psi_{h,n,p}(\tau_1,\tau_2,x) $ is some 
antisymmetric eigenfunctions in time which $n$ labels the fourier mode for the sum of the two times, and $h$ specifies the dependence on the difference of the times, and $p$ labels the fourier mode for space. Technically, one can further simplify the calculation using the symmetrized kernel\footnote{Roughly speaking, this trick is used to avoid the computation of the inner product $\langle \Psi | F_0 \rangle$}\cite{maldacena2016comments} :
\begin{equation}
\tilde{K}_{xy}(\tau_1,\tau_2,\tau_3,\tau_4):=
\begin{tikzpicture}[baseline={([yshift=-4pt]current bounding box.center)}]
\filldraw[fill=black] (-40pt,20pt) circle (1pt) node[left] {$1,x$};
\draw[thick] (-40pt,20pt)..controls (-30pt,10pt) and (-30pt,-10pt)..(-40pt,-20pt);
\filldraw[fill=black] (10pt,20pt) circle (1pt) node[right] {$3,y$};
\filldraw[fill=black] (-40pt,-20pt) circle (1pt) node[left] {$2,x$};
\filldraw[fill=black] (10pt,-20pt) circle (1pt) node[right] {$4,y$};
\draw[thick] (10pt,-20pt)--(-40pt,-20pt);
\draw[thick] (-40pt,20pt)--(10pt,20pt);
\draw[thick] (10pt,20pt)..controls (0pt,10pt) and (0pt,-10pt)..(10pt,-20pt);
\end{tikzpicture}=S_{xy}\tilde{K}(\tau_1,\tau_2,\tau_3,\tau_4)
\end{equation}
where the symmetrized temporal kernel is defined as
$
\tilde{K}(\tau_1,\tau_2,\tau_3,\tau_4)=
3\Jeff^2 G^s(\tau_{13})\cdot |G^s(\tau_{34})|\cdot G^s(\tau_{42}) \cdot | G^s(\tau_{21})|$. 
The simplification works by the following steps: (1) add two rungs (with absolute value for convenience) to the ladders in $\calF$ (one on the left\footnote{the extra factor $3J^2 S_{xx'}$ is needed to construct the kernel}, one on the right), see figure~\ref{fig: diagrams using symmetrized kernel};
\begin{figure}
[t]
\center
\[
\begin{tikzpicture}[scale=0.8, baseline={([yshift=-4pt]current bounding box.center)}]
\draw[thick] (-40pt,20pt)..controls (-30pt,10pt) and (-30pt,-10pt)..(-40pt,-20pt);
\end{tikzpicture}
~
\cdot
~
\begin{tikzpicture}[scale=0.8,baseline={([yshift=-4pt]current bounding box.center)}]
\draw[thick] (10pt,-20pt)--(-40pt,-20pt);
\draw[thick] (-40pt,20pt)--(10pt,20pt);
\draw[thick] (10pt,20pt)..controls (0pt,10pt) and (0pt,-10pt)..(10pt,-20pt);
\draw[thick] (10pt,20pt)..controls (20pt,10pt) and (20pt,-10pt)..(10pt,-20pt);
\draw[thick] (60pt,-20pt)--(10pt,-20pt);
\draw[thick] (10pt,20pt)--(60pt,20pt);
\draw[thick] (60pt,20pt)..controls (50pt,10pt) and (50pt,-10pt)..(60pt,-20pt);
\node at(35pt,0pt){$\ldots$};
\draw[thick] (60pt,20pt)..controls (70pt,10pt) and (70pt,-10pt)..(60pt,-20pt);
\draw[thick] (110pt,-20pt)--(60pt,-20pt);
\draw[thick] (60pt,20pt)--(110pt,20pt);
\end{tikzpicture}
~
\cdot
~
\begin{tikzpicture}[scale=0.8,baseline={([yshift=-4pt]current bounding box.center)}]
\draw[thick] (110pt,20pt)..controls (100pt,10pt) and (100pt,-10pt)..(110pt,-20pt);
\end{tikzpicture}
=
\begin{tikzpicture}[scale=0.8,baseline={([yshift=-4pt]current bounding box.center)}]
\draw[thick] (-40pt,20pt)..controls (-30pt,10pt) and (-30pt,-10pt)..(-40pt,-20pt);
\draw[thick] (10pt,-20pt)--(-40pt,-20pt);
\draw[thick] (-40pt,20pt)--(10pt,20pt);
\draw[thick] (10pt,20pt)..controls (0pt,10pt) and (0pt,-10pt)..(10pt,-20pt);
\draw[thick] (10pt,20pt)..controls (20pt,10pt) and (20pt,-10pt)..(10pt,-20pt);
\draw[thick] (60pt,-20pt)--(10pt,-20pt);
\draw[thick] (10pt,20pt)--(60pt,20pt);
\draw[thick] (60pt,20pt)..controls (50pt,10pt) and (50pt,-10pt)..(60pt,-20pt);
\node at(35pt,0pt){$\ldots$};
\draw[thick] (60pt,20pt)..controls (70pt,10pt) and (70pt,-10pt)..(60pt,-20pt);
\draw[thick] (110pt,-20pt)--(60pt,-20pt);
\draw[thick] (60pt,20pt)--(110pt,20pt);
\draw[thick] (110pt,20pt)..controls (100pt,10pt) and (100pt,-10pt)..(110pt,-20pt);
\end{tikzpicture}\]
\caption{A trick\cite{maldacena2016comments} to simplify the expressions by using the symmetrized kernel.}
\label{fig: diagrams using symmetrized kernel}
\end{figure}
(2) we get a multiple of ``curved boxes''
, then express it as a power of the symmetrized kernel; (3) in the end, sum over all ladders, which is now a geometric series:
\begin{align}
\sum_{x'} 3J^2 S_{xx'}|G^s(\tau_{12})| \calF_{x'y}(\tau_1,\tau_2;\tau_3,\tau_4)  |G^s(\tau_{34}|
=2 \sum_{n=1}^\infty
\begin{tikzpicture}[scale=0.5,baseline={([yshift=-4pt]current bounding box.center)}]
\filldraw[fill=black] (-40pt,20pt) circle (1pt) node[left] {$1$};
\filldraw[fill=black] (110pt,20pt) circle (1pt) node[right] {$3$};
\filldraw[fill=black] (-40pt,-20pt) circle (1pt) node[left] {$2$};
\filldraw[fill=black] (110pt,-20pt) circle (1pt) node[right] {$4$};
\draw[thick] (-40pt,20pt)..controls (-30pt,10pt) and (-30pt,-10pt)..(-40pt,-20pt);
\draw[thick] (10pt,-20pt)--(-40pt,-20pt);
\draw[thick] (-40pt,20pt)--(10pt,20pt);
\draw[thick] (10pt,20pt)..controls (0pt,10pt) and (0pt,-10pt)..(10pt,-20pt);
\draw[thick] (10pt,20pt)..controls (20pt,10pt) and (20pt,-10pt)..(10pt,-20pt);
\draw[thick] (60pt,-20pt)--(10pt,-20pt);
\draw[thick] (10pt,20pt)--(60pt,20pt);
\draw[thick] (60pt,20pt)..controls (50pt,10pt) and (50pt,-10pt)..(60pt,-20pt);
\node at(35pt,0pt){$\ldots$};
\draw[thick] (60pt,20pt)..controls (70pt,10pt) and (70pt,-10pt)..(60pt,-20pt);
\draw[thick] (110pt,-20pt)--(60pt,-20pt);
\draw[thick] (60pt,20pt)--(110pt,20pt);
\draw[thick] (110pt,20pt)..controls (100pt,10pt) and (100pt,-10pt)..(110pt,-20pt);
\end{tikzpicture}
=2 \sum_{n=1}^\infty (\tilde{K}^n)_{xy} 
\end{align}
where the factor of $2$ comes from the counting for extra term with $3\leftrightarrow 4$. 
This is the general formal expression for the connected Euclidean four-point function. We can proceed by diagonalizing the spatial kernel via plane waves $e^{ipx}$, i.e., $k(h,n,p)=s(p) k(h,n)$ and write the general expression above in momentum space:
\begin{align}
\calF_p(\tau_1,\tau_2;\tau_3,\tau_4) = \frac{2}{3\Jeff^2 |G^s(\tau_{12})| \cdot| G^s(\tau_{34})|}  \sum_{h,n} \frac{k(h,n)}{1-s(p) k(h,n)} \Psi_{h,n}(\tau_1,\tau_2)   \Psi^*_{h,n}(\tau_3,\tau_4)
\end{align}
where $k(h,n)$ is the eigenvalue of temporal kernel. This agrees with equation~\ref{eqn: general formula for finite $p$} derived from effective action.

\section{Summation trick and the prefactor}
\label{appendix: summation trick}

In this appendix we determine the large $t$ behavior of the contribution to the OTOC from the $h = 2$ modes. We choose a special configuration:
\begin{equation}
\tau_{12}=\tau_{34}=\beta/2,\quad  |\Re(\tau)|<\pi/2
\end{equation}
where $\tau= \frac{\tau_1+\tau_2-\tau_3-\tau_4}{2}$ is the center of mass time separation, and the requirement $|\Re(\tau)|<\beta/4$ ensures the out-of-time order. The sum over $h = 2$ modes in equation~(\ref{eqn: h=2 finite momentum})  is
\begin{equation}
\frac{\calF_{p,h=2}(\tau)}{G^s(\frac{\beta}{2})^2} = \frac{32\Jeff}{\sqrt{2} \alpha_K} \sum_{n\geq 2~even} \frac{(-1)^{n/2} \cos (n \frac{2\pi}{\beta} \tau) }{  n^2-1  } \frac{n}{\frac{2\pi n}{\beta}+Dp^2}.
\end{equation}
We would like to continue $\tau = it$ and determine the large $t$ behavior of this sum. To do this we consider the following integral
\begin{equation}
I=
\frac{1}{2\pi i} \int_{-i\infty + 0}^{i\infty+0} d\omega~ \frac{\pi}{2} \cdot \frac{\cos ( \omega \frac{2\pi}{\beta} \tau) }{\sin (\pi \omega /2)}\cdot  \frac{ \omega }{\omega ^2-1} \cdot \frac{1}{\frac{2\pi \omega }{\beta}+Dp^2}
\end{equation}
Convergence is guaranteed by $|\Re(\tau)| < \beta/4$. The integrand has poles at $\omega =1$ and $2,4,6.\ldots$ on $\RR^+$.
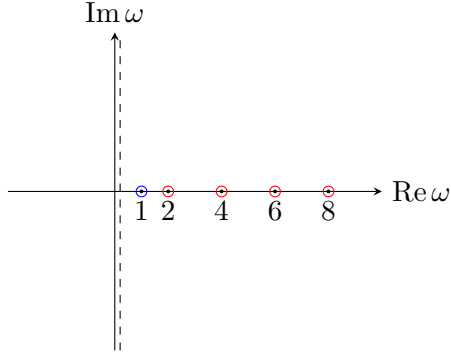
\begin{figure}
[h]
\center
\begin{tikzpicture}[baseline={(current bounding box.center)}]
\draw[->,>=stealth] (-40pt,0pt)--(100pt,0pt) node [right] {$\Re \omega$};
\draw[->,>=stealth] (0pt,-60pt)--(0pt,60pt) node [above] {$\Im \omega$};
\draw[dashed] (2pt,-60pt)--(2pt,60pt) ;
\filldraw[fill=black] (10pt,0pt) circle (0.5pt) node[below]{$1$};
\filldraw[fill=black] (20pt,0pt) circle (0.5pt) node[below]{$2$};
\filldraw[fill=black] (40pt,0pt) circle (0.5pt) node[below]{$4$};
\filldraw[fill=black] (60pt,0pt) circle (0.5pt) node[below]{$6$};
\filldraw[fill=black] (80pt,0pt) circle (0.5pt) node[below]{$8$};
\draw[blue] (10pt,0pt) circle (2pt) ;
\draw[red]  (20pt,0pt) circle (2pt) ;
\draw[red]  (40pt,0pt) circle (2pt) ;
\draw[red]  (60pt,0pt) circle (2pt) ;
\draw[red] (80pt,0pt) circle (2pt) ;
\end{tikzpicture}
\caption{Contour deformation from dashed line to the blue and red circles surrounding the poles on the positive real axis.}
\label{fig: poles in summation trick}
\end{figure}
 When we deform the contour to the right  (see figure~\ref{fig: poles in summation trick}), we find
\begin{equation}
I= -\sum_{n\geq 2~even} \frac{(-1)^{n/2} \cos (n \frac{2\pi}{\beta}\tau) }{  n^2-1  } \frac{n}{\frac{2\pi n}{\beta}+Dp^2 } - \frac{\pi}{4} \frac{\cos (\frac{2\pi}{\beta}\tau)}{\frac{2\pi}{\beta}+Dp^2}.
\end{equation} 
The key point is that when we continue to large real time $\tau = it$, the integral $I$ remains convergent and non-growing in time, so we must have
\begin{equation}
\sum_{n\geq 2~even} \frac{(-1)^{n/2} \cos (n \frac{2\pi}{\beta}\tau) }{  n^2-1  } \frac{n}{\frac{2\pi n}{\beta}+Dp^2 }  = - \frac{\pi}{4} \frac{\cos (\frac{2\pi}{\beta}\tau)}{\frac{2\pi}{\beta}+Dp^2} + \text{(non-growing)}.
\end{equation}
This directly gives (\ref{h=2chaos}).

\section{Diffusion and the butterfly velocity in general dimensions}
\label{appendix: higher dim}

In this section, we sketch the computation relevant to the diffusion and the butterfly velocity in general dimensional models with translation symmetry. We won't derive the exact formula for most general case, but will instead present key steps that determines the diffusion constant and butterfly velocity.

In the mode with transnational symmetry, the four-point function has simple expression in terms of momentum eigenvalue:
\begin{equation}
\calF_{\vec{p}(}\tau_1,\tau_2;\tau_3,\tau_4) = \frac{2}{3J^2 G^s(\tau_{12})G^s(\tau_{34})} \sum_{h,n} \frac{k(h,n)}{1-k(h,n)s(\vec{p})} \Psi_{h,n}(\tau_1,\tau_2) \Psi^*_{h,n}(\tau_3,\tau_4)
\end{equation}
Notice here the momentum $\vec{p}$ represents a general high dimensional vector. Further restrict the model to be local, we have a small $p$ expansion for eigen-value $s(\vec{p})\simeq 1-\sum_j a_j p_j^2$.

As we have noticed in the SYK and SYK chain model, the contribution relevant to the energy fluctuation arises from $h=2$ modes, which corresponds to the reparametrization fields (pseudo-Goldstone modes). At strongly coupling limit, the eigenvalue for $h=2$ receives a correction depends on $n$. $k(h=2,n)\simeq 1-\frac{\sqrt{2} \alpha_K}{\beta J} |n|+ \ldots$. Therefore, the pole that determines the diffusion constant has the simple form:
\begin{align}
\calF_{\vec{p}} &\propto \sum_n \frac{1}{1-(1-\frac{\sqrt{2} \alpha_K |n|}{\beta J}) (1-\sum_j a_j p_j^2)}+ \ldots \nn \\
& \propto  \sum_n \frac{1}{\frac{\sqrt{2} |n| \alpha_K}{\beta J} +\sum_j a_j p_j^2}+ \ldots 
\end{align}
this is the formula in Matsubara frequency $\omega_n=\frac{2\pi}{\beta} n$, after rotating to real time, we have diffusion pole:
\begin{equation}
\frac{1}{-i\omega + \sum_j D_j p_j^2},\quad D_j = \frac{ 2\pi \alpha_j J}{\sqrt{2} \alpha_K}
\end{equation}

Next is to compute the butterfly velocity, which can be extract from the OTO correlation function. Again, we choose the time configuration which placed four time equally spacing around imaginary time circle, i.e., we are computing:
\begin{eqnarray}
F(\vec{x},t)&=&\frac{1}{N^2}\sum_{j,k=1}^N \left\langle\chi_{j,\vec{x}} (t+i\frac{3\beta}4) \chi_{k,0}(i\frac{\beta}2)  \chi_{j,\vec{x}}(t+i\frac{\beta}4) \chi_{k,0}(0) \right\rangle_\beta
\nonumber\\
&=&\frac{1}{N^2}\sum_{j,k=1}^N  \Tr \left( r \chi_{j,\vec{x}} (t) r \chi_{k,0}(0) r \chi_{j,\vec{x}}(t') r \chi_{k,0}(0) \right)
\end{eqnarray}
Here the spatial coordinate $\vec{x}$ represents a high dimensional vector. Analogous to the computation in the chain model, we goes to momentum space, and plug in the $h=2$ eigen-functions, which is the leading contribution:
\begin{equation}
F(\vec{p},t)_{h=2} \propto \sum_{n\geq 2~even} \frac{(-1)^{n/2} \cos (n \frac{2\pi}{\beta} \tau) }{  n^2-1  } \frac{n}{\frac{2\pi n}{\beta}+\sum_j D_jp_j^2} 
\end{equation}
Using the trick in appendix~\ref{appendix: summation trick} and analytic to real time, we have a formula in momentum space:
\begin{equation}
F(\vec{p},t)_{h=2} \propto \frac{1}{\frac{2\pi }{\beta}+\sum_j D_jp_j^2} e^{\frac{2\pi}{\beta} t}
\end{equation}
Here we get a Lyapunov exponent $\lambda_L$ from $h=2$ contribution. In general, the exponent will receive a $1/\beta J$ correction as well as small $p^2$ correction from $h\neq 2$ part. Following the argument in section~\ref{section: chaos and the butterfly velocity}, we have
\begin{equation}
F(\vec{p},t) \propto \frac{1}{\frac{2\pi }{\beta}+\sum_j D_jp_j^2} \exp\left\lbrace \frac{2\pi}{\beta} \left[ 1- c \left( \frac{2\pi }{\beta}+\sum_j D_jp_j^2 \right)   \right] t \right\rbrace,\quad c=\frac{3\alpha_K}{2\sqrt{2} \pi  J}
\end{equation}
In general dimension, we need to compute the following fourier transformation to get the butterfly velocity in $x_j$ direction:
\begin{equation}
F(x_j,t) \propto \int \frac{	d^d p}{(2\pi)^d} \frac{e^{ i p_j x_j}}{\frac{2\pi }{\beta}+\sum_j D_jp_j^2} \exp\left\lbrace \frac{2\pi}{\beta} \left[ 1- c \left( \frac{2\pi }{\beta}+\sum_j D_jp_j^2 \right)   \right] t \right\rbrace
\end{equation}
Notice $t\sim \beta \log N$ is a large parameter. Therefore, 
we can integral over all direction except $p_j$ by saddle point approximation, where the saddle point is $p_k=0$ $k\neq j$:
\begin{equation}
F(x_j,t) \propto t^{-(d-1)/2}\int \frac{	dp_j}{(2\pi)^d} \frac{e^{ i p_j x_j}}{\frac{2\pi }{\beta}+ D_jp_j^2} \exp\left\lbrace \frac{2\pi}{\beta} \left[ 1- c \left( \frac{2\pi }{\beta}+D_jp_j^2 \right)   \right] t \right\rbrace
\end{equation}
Then this fourier transformation essentially goes back to the 1-dimensional case, for which we know that at large $x_j$, the integral is dominated by the pole $\frac{2\pi}{\beta} + D_j p_j^2=0$. The pole determines the exponential decaying profile for $F(x_j,t)$ at real space, \begin{equation}
F(x_j,t) \propto \exp\left[ \frac{2\pi}{\beta} (t - |x_j|/v_{B,j}) \right]
\end{equation}
with butterfly velocity 
\begin{equation}
v_{B,j}^2=   2\pi T D_j
\end{equation}

\bibliography{ref}

\end{document}